\documentclass[12pt, draftclsnofoot, onecolumn]{IEEEtran}
\usepackage{amsmath,amssymb}
\usepackage[dvips]{graphicx}
\usepackage{amsfonts}
\usepackage[mathscr]{eucal}
\usepackage{latexsym}
\usepackage{amsthm}
\usepackage{exscale}
\usepackage[mathscr]{eucal}
\usepackage{bm}
\usepackage[dvipsnames]{color}
\usepackage{cases}
\usepackage{color}
\usepackage{epsfig}
\usepackage{algorithm}
\usepackage{algorithmic}
\usepackage[verbose,nospace,sort]{cite}
\usepackage{tabularx}
\usepackage{multirow}
\usepackage{multicol}
\usepackage{balance}
\usepackage{caption}
\usepackage{subcaption}
\usepackage{setspace}

\scrollmode

\theoremstyle{definition}

\newtheorem{remark}{Remark}
\newtheorem{lemma}{Lemma}

\newcommand{\q}{\mathbf q}

\newcommand{\me}{\mathrm{me}}
\newcommand{\mr}{\mathrm{mr}}
\newcommand{\tr}{\mathrm{tr}}
\newcommand{\hov}{\mathrm{hov}}
\newcommand{\tot}{\mathrm{tot}}
\newcommand{\hc}{\mathrm{hc}}

\graphicspath{{./figs/}}

\begin{document}
\title{Energy Minimization for Wireless Communication with Rotary-Wing UAV}
\author{Yong Zeng, {\it Member,~IEEE,} Jie Xu, {\it Member,~IEEE,} and Rui Zhang, {\it Fellow,~IEEE} \vspace{-5ex}  \\
\thanks{Y. Zeng and R. Zhang are with the Department of Electrical and Computer Engineering, National University of Singapore (e-mail: \{elezeng, elezhang\}@nus.edu.sg).}
\thanks{J. Xu is with the School of Information Engineering, Guangdong University of Technology (e-mail: jiexu@gdut.edu.cn).}
}

\maketitle

\begin{abstract}
This paper studies unmanned aerial vehicle (UAV) enabled  wireless communication, where a rotary-wing UAV is dispatched to send/collect data to/from multiple ground nodes (GNs). We aim to minimize the total UAV energy consumption, including both propulsion energy and communication related energy, while satisfying the communication throughput requirement of each GN. To this end, we first derive an analytical propulsion power consumption model for rotary-wing UAVs, and then formulate the energy minimization problem by jointly optimizing the UAV trajectory and communication time allocation among GNs, as well as the total mission completion time. The problem is difficult to be optimally solved, as it is non-convex and involves infinitely many variables over time. To tackle this problem,  we first consider the simple {\it fly-hover-communicate} design, where the UAV successively visits a set of hovering locations and communicates with one corresponding GN when  hovering at each location. For this design, we propose an efficient algorithm to optimize the hovering locations and durations, as well as the flying trajectory connecting these hovering locations, by leveraging the travelling salesman problem (TSP) and convex optimization techniques. Next, we consider the  general case where the UAV  communicates also when flying. We  propose a new {\it path discretization} method to transform the original problem into a discretized equivalent with a finite number of optimization variables, for which we obtain a locally optimal solution by applying the successive convex approximation (SCA) technique. Numerical results show the significant performance gains of the proposed designs over benchmark schemes, in achieving energy-efficient communication with rotary-wing UAVs.
\end{abstract}


\begin{center}
{ \bf Index Terms}
\end{center}
{\small
UAV communication, rotary-wing UAV, energy  model, energy-efficient communication,  trajectory optimization, path discretization.
}

\section{Introduction}
Wireless communication using unmanned aerial platforms is a promising technology to achieve wireless coverage in areas without or with insufficient terrestrial infrastructures. Early efforts have been primarily focusing on using high altitude platforms (HAPs), which are deployed in stratosphere at altitude around 20 km, aiming to provide ubiquitous coverage in rural or remote areas. These include the Project Loon by Google with the mission of ``Balloon-powered Internet for everyone'', as well as the Project Skybender by Google and the Project Aquila by Facebook, both using solar-powered drones to provide internet access from the sky. On the other hand, wireless communication using low altitude platforms (LAPs), typically below a few kilometers above the ground, has received growing interests recently. LAPs can be implemented in various ways, such as helikite \cite{912} and unmanned aerial vehicles (UAVs) \cite{649,949,913,952,967}. In particular, compared to other airborne solutions such as HAPs and helikite, UAV-enabled wireless communication brings new advantages \cite{649}, such as on-demand and more swift deployment, superior link quality in the presence of shorter-distance line-of-sight (LoS) communication channel with ground nodes (GNs), and higher network flexibility with the fully controllable UAV movement in three dimensional (3D) airspace. Therefore, UAV-enabled wireless communication has many potential use cases, including public safety communication, temporary traffic offloading for cellular base stations (BSs), information dissemination and data collection for Internet of Things (IoTs), as well as emergency response and fast service recovery after natural disasters.

Prior researches on UAV-enabled wireless  communications can be loosely classified into two categories. In the first category, UAVs are deployed as (quasi-)stationary aerial BSs. In this case, UAVs resemble the conventional static terrestrial BSs, but at a much higher altitude and thus possesses new channel characteristics \cite{971,621,962,963,964}. In particular, it was shown that as the UAV altitude increases, the LoS probability between the UAV and GNs also increases \cite{642}. By exploiting such unique channel characteristics, significant efforts have been devoted to studying the various aspects of UAV-enabled BSs, such as UAV placement optimization \cite{642,803,886,914,922,940}, performance analysis \cite{974,954,966}, spectrum sharing \cite{955}, and cell association \cite{965}. 
In contrast, the other category considers the application scenarios where UAVs are employed as mobile BSs/relays/access points (APs) \cite{641,887,904,918,919,957}, whose trajectories can be designed to optimize the communication performance. For example, a UAV as a mobile relay or data collector can fly closer to its associated GNs for communication to improve the overall spectrum efficiency \cite{641} and/or save the communication energy of GNs \cite{918}. In \cite{641}, a new framework of joint power allocation and UAV trajectory optimization was proposed for the UAV-enabled mobile relaying system, which has been extended to various other setups such as  UAV-enabled data collection \cite{918}, multi-UAV coordinated/cooperative communication \cite{919}, \cite{975},  and UAV-enabled wireless power transfer \cite{956}.

One critical issue of UAV-enabled wireless communication lies in the limited on-board energy of UAVs \cite{649}, which needs to be efficiently used to enhance the communication performance and prolong the UAV's endurance. Compared to conventional terrestrial BSs, UAVs incur additional  propulsion energy consumption to maintain airborne and support their movement. In practice, the UAV propulsion power is usually much higher than the communication related power. As a result, the energy-efficient wireless communication design with UAV is significantly different from that in conventional terrestrial communication systems. An initial attempt for designing energy-efficient UAV communication via trajectory optimization was made in \cite{904}, where the energy efficiency in bits/Joule of a {\it fixed-wing} UAV enabled communication system is maximized for a given flight duration. To that end, a generic energy model as a function of the UAV's velocity and acceleration was derived for fixed-wing UAVs. Based on the energy model in \cite{904}, the authors in \cite{960} further revealed an interesting trade-off between UAV's energy consumption and that of the GNs it communicating with. However, the above results for fixed-wing UAVs cannot be applied for {\it rotary-wing} UAVs, due to their fundamentally different mechanical designs and hence drastically different propulsion energy models. This thus motivates our current work to investigate energy-efficient communication for rotary-wing UAVs.


In this paper, we study a wireless communication system enabled a rotary-wing UAV. Compared to fixed-wing UAVs, rotary-wing UAVs have several appealing advantages such as the ability to take off and land vertically, as well as for hovering, which render them more popular in the current UAV market. We consider the scenario where a rotary-wing UAV is dispatched as a flying AP to communicate with multiple GNs, each of which has a target number of information bits to be transmitted/received to/from the UAV. Such a setup corresponds to many practical applications, such as UAV-enabled data collection for periodic sensing, UAV-enabled caching where the UAV pre-fetches the data and then transmits to the designated caching nodes \cite{958}, etc. Our objective is to minimize the UAV's energy consumption, including both propulsion energy and communication energy, while ensuring that the communication requirement for each GN is satisfied. The main contributions of this paper are summarized as follows.

First, we  derive an analytical model for the propulsion power consumption of rotary-wing UAVs, based on the results in aircraft literature \cite{905}, \cite{766}. As expected, the obtained model is significantly different from that for fixed-wing UAVs derived in our prior work \cite{904}.

 Based on the derived power consumption model, we formulate the energy minimization problem that jointly optimizes the UAV trajectory, the communication time allocation among the multiple GNs, as well as the total mission completion time. The problem is difficult to be optimally solved, as it is non-convex and constitutes infinite number of optimization variables that are coupled in continuous functions over time. To tackle this problem, we first consider the simple {\it fly-hover-communicate} design \cite{940} to gain useful insights. Under this design, the UAV successively visits a set of optimized hovering locations, and  communicates with each of the GNs only when hovering at the corresponding location. In this case, the problem reduces to finding the optimal hovering locations and hovering duration at each location, as well as the visiting order and flying speed among these locations. The problem is still NP hard, as it includes the classic NP hard travelling salesman problem (TSP) as a special case \cite{TSP}. By leveraging the existing TSP-solving algorithm \cite{908} and convex optimization techniques, an efficient high-quality approximate solution is obtained for our problem.

Next, we propose a general solution to the energy minimization problem where the UAV communicates also when flying. To this end, we first propose a novel discretization technique, called {\it path discretization}, to transform the original problem with infinitely many variables into a more tractable form with a finite number of variables. Different from the widely used {\it time discretization} approach for UAV trajectory design (see e.g. \cite{641} and \cite{904}), path discretization does not require the mission completion time to be pre-specified. This is particularly useful for problems where the mission completion time  is also one of the optimization variables, as for the energy minimization problem studied in this paper. However, the path-discretized problem is still non-convex, and thus it is challenging to find its optimal solution. By utilizing the successive convex approximation (SCA) technique \cite{641}, an efficient iterative algorithm is proposed to simultaneously update the UAV trajectory and communication time allocation at each iteration, which is guaranteed to converge to at least a locally optimal solution satisfying the  Karush-Kuhn-Tucker (KKT) conditions. Last, simulation results are provided to validate the proposed designs and show their significant performance gains over benchmark schemes.

It is worth noting that another related line of work is on mobile robotics, which exploits the mobility of ground robots for various applications \cite{969}, \cite{970}. However, the design for UAV communication systems are significantly different from that for ground robotics due to the distinct air-to-ground channel characteristics \cite{971,621,962,963,964} as well as the fundamentally different energy consumption models. For example, the power consumption of mobile robots can usually be modeled as a polynomial and monotonically increasing function with respect to its moving speed \cite{969}, which is much simpler than that for fixed-wing UAVs as in \cite{904} and rotary-wing UAVs in Section~\ref{sec:model} of the current work.

\section{System Model and Problem Formulation}\label{sec:SystemModel}
\subsection{System Model}
We consider a wireless communication system where a rotary-wing UAV is dispatched to communicate with $K$ GNs, which are denoted by the set $\mathcal K=\{1,\cdots, K\}$. The horizontal location of the GN $k\in \mathcal K$ is denoted as $\mathbf w_k\in \mathbb{R}^{2\times 1}$. 
 We assume that the UAV flies at a constant altitude $H$ and the total number of information bits that need to be communicated with GN $k$ is  $\tilde Q_k$. Let $T_t$ denote the total time required for the UAV to complete the mission, which is a design variable. 
Denote by $\mathbf q(t)\in \mathbb{R}^{2\times 1}$ with $0\leq t \leq T_t$ the UAV trajectory projected onto the horizontal plane. Let $V_{\max}$ denote the maximum UAV speed. We then have the constraint $\|\dot{\mathbf q}(t)\|\leq V_{\max}$. At any time instant $t\in [0, T_t]$,  the distance between the UAV and  GN $k$ is given by $d_k(t)=\sqrt{H^2+\|\mathbf q(t)-\mathbf w_k\|^2}, \ k\in \mathcal K$.

We assume that the wireless channels between the UAV and GNs are dominated by LoS links. Thus, the channel power gain between the UAV and GT $k$ can be modeled based on the free-space path loss model as
\begin{align}
h_k(t) = \beta_0 d_k^{-2}(t) = \frac{\beta_0}{H^2	+\|\mathbf q(t)-\mathbf w_k\|^2},
\end{align}
where $\beta_0$ represents the channel power gain at the reference distance of $1$ meter (m). Furthermore, assuming a fixed transmission power $P$ by the transmitter when it is scheduled for communication, the achievable rate in bits per second (bps) between GN $k$ and the UAV at any time instant $t$ is expressed as
\begin{align}
R_k(t)&=B \log_2\left(1+ \frac{Ph_k(t)}{\sigma^2 \Gamma}\right)=B \log_2\left(1+ \frac{\gamma_0}{H^2+\|\mathbf q(t)-\mathbf w_k\|^2}\right), \label{eq:Rkt}
\end{align}
where $B$ denotes the channel bandwidth in hertz (Hz), $\sigma^2$ is the noise power at the receiver, $\Gamma>1$  accounts for the gap from the channel capacity due to  the practical modulation and coding scheme employed, and $\gamma_0\triangleq P\beta_0/(\sigma^2\Gamma)$ is defined as the received signal-to-noise ratio (SNR) at the reference distance of $1$ m.

We assume that the time-division multiple access (TDMA) protocol is applied for the UAV to serve the $K$ GNs, in order to fully exploit the time-varying channels with trajectory design. Let a binary variable $\lambda_k(t)\in \{0, 1\}$ denote the user scheduling indicator at time instant $t$,  with $\lambda_k(t)=1$ indicating that GN $k$ is scheduled for communication at instant $t$ and $\lambda_k(t)=0$ otherwise. As at most one GN can be scheduled at each time instant $t$,  we have
\begin{align}
\sum_{k=1}^K \lambda_k(t)\leq 1, \forall t\in [0, T_t]. \label{eq:lambdakt}
\end{align}

Therefore, the aggregated communication throughput for GN $k$ is a function of $T_t$, $\mathbf q(t)$, and $\lambda_k(t)$, which can be expressed as
\begin{align}\label{eq:Rbark}
\bar R_k&\big(T_t, \{\mathbf q(t)\}, \{\lambda_k(t)\}\big) = \int_0^{T_t} \lambda_k(t) R_k(t)dt\notag \\
&=  B \int_0^{T_t} \lambda_k(t) \log_2\left(1+ \frac{\gamma_0}{H^2+\|\mathbf q(t)-\mathbf w_k\|^2}\right)dt.
\end{align}
To ensure the target communication throughput requirement for each GN $k$, we must have
\begin{align}
\bar R_k\big(T_t, \{\mathbf q(t)\}, \{\lambda_k(t)\}\big)\geq \tilde {Q}_k, \ \forall k\in \mathcal K. \label{eq:ThroughputReq}
\end{align}


\subsection{Energy Consumption Model for Rotary-Wing UAV}\label{sec:model}
The UAV energy consumption is in general composed of two main components, namely the communication related energy and the propulsion energy. The communication related energy includes that for communication circuitry, signal processing, signal radiation/reception, etc. In this paper, we assume that the communication related power is a constant, which is denoted as $P_c$ in watt (W).  On the other hand, the propulsion energy consumption is needed to keep the UAV aloft and support its movement, if necessary. In general, the propulsion energy depends on the UAV flying speed as well as its acceleration. In this paper, for the purpose of exposition and drawing the essential design insight, we ignore the additional energy consumption caused by UAV acceleration, which is valid for typical communication applications where UAV manoeuvring time only takes a small portion of the total operation time. As derived in Appendix~\ref{A:energyModel}, for a rotary-wing UAV flying with speed $V$, the propulsion power consumption  can be modeled as
\begin{align}
P(V)=& \underbrace{P_0 \left(1 + \frac{3 V^2}{U_{\mathrm{tip}}^2} \right)}_{\text{blade profile}}+ \underbrace{P_i  \left( \sqrt{1 + \frac{V^4}{4 v_0^4}}-\frac{V^2}{2v_0^2}\right)^{1/2}}_{\text{induced}}+ \underbrace{\frac{1}{2} d_0 \rho s A V^3}_{\text{parasite}},\label{eq:Pstr}
\end{align}
where $P_0$ and $P_i$ are two constants defined in \eqref{eq:Phover} of Appendix~\ref{A:energyModel} representing the {\it blade profile power} and {\it induced power} in hovering status, respectively, $U_{\mathrm{tip}}$ denotes the tip speed of the rotor blade, $v_0$ is known as the mean rotor induced velocity in hover, $d_0$ and $s$ are the fuselage drag ratio and rotor solidity, respectively, and $\rho$ and $A$ denote the air density and rotor disc area, respectively. The relevant parameters are explained in details in Table~\ref{table:notations} and Appendix~\ref{A:energyModel}. It is observed from \eqref{eq:Pstr} that the propulsion power consumption of rotary-wing UAVs consists of three components: blade profile, induced, and parasite power. The blade profile power and parasite power, which increase quadratically and cubically with $V$, respectively, are needed to overcome the profile drag of the blades and the fuselage drag, respectively. On the other hand, the induced power is that required to overcome the induced drag of the blades, which decreases with $V$.

By substituting $V=0$ into \eqref{eq:Pstr}, we obtain the power consumption for hovering status as $P_h=P_0+P_i$, which is a finite value depending on the aircraft weight, air density, and rotor disc area, etc. (see \eqref{eq:Phover} in Appendix~\ref{A:energyModel} for details). As $V$ increases, it can be verified that $P(V)$ in \eqref{eq:Pstr} firstly decreases and then increases with $V$, i.e., hovering is in general not the most power-conserving status. It can be verified that the power function $P(V)$ in \eqref{eq:Pstr} is neither convex nor concave with respect to $V$. It is much more involved compared to the power model for fixed-wing UAV (cf. Equation (7) of \cite{904}), which is a convex function consisting of two simple terms: one increasing cubically and the other decreasing inversely with $V$.

\begin{figure}
\centering
\includegraphics[scale=0.4]{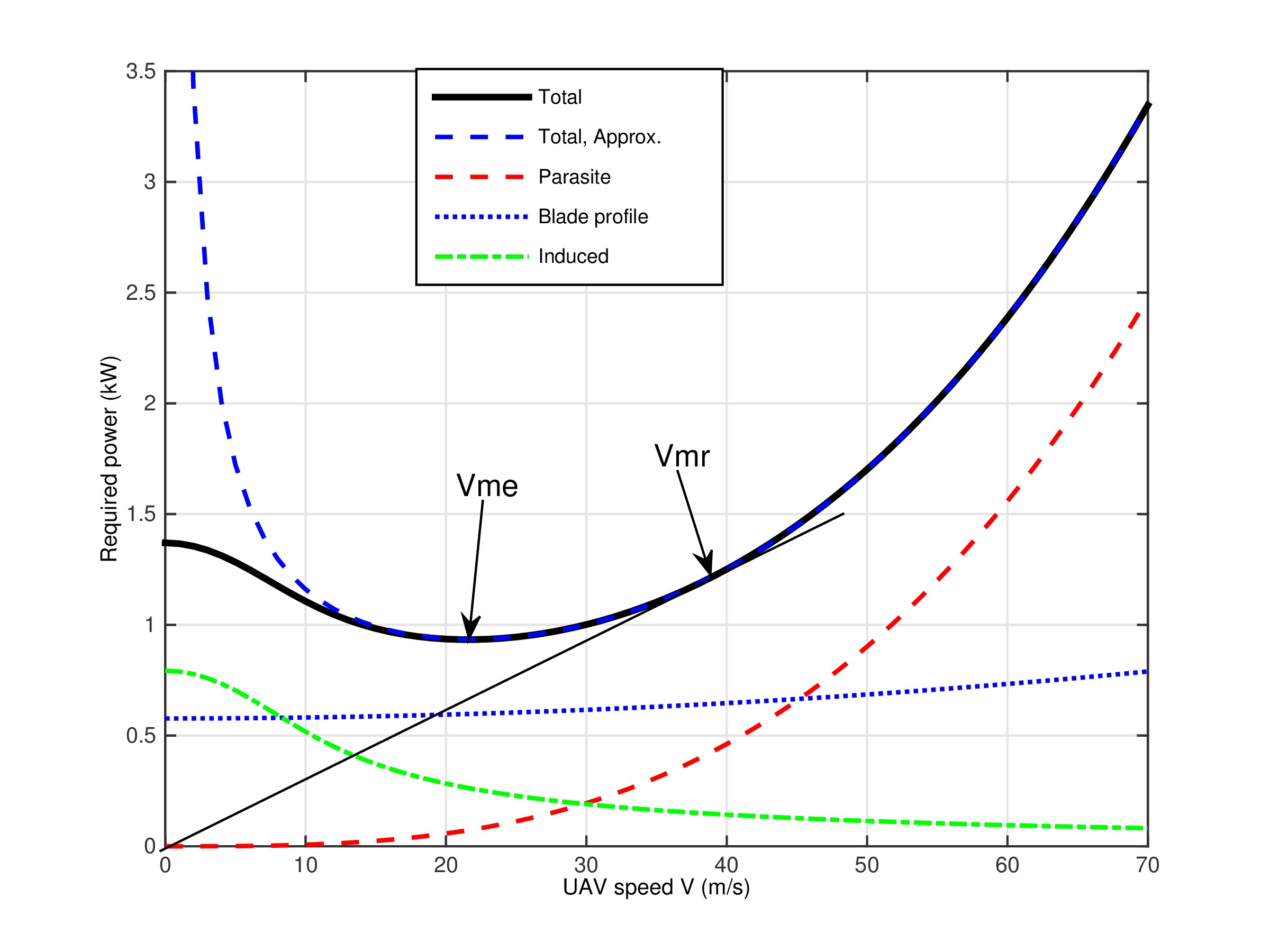}
\caption{Propulsion power consumption versus speed $V$ for rotary-wing UAV.\vspace{-5ex}}\label{F:PowervsSpeed}
\end{figure}

When $V\gg v_0$,  by applying the first-order Taylor approximation $(1+x)^{1/2}\approx 1+\frac{1}{2}x$ for $|x|\ll 1$, \eqref{eq:Pstr} can be approximated as a convex function, i.e.,
\begin{align}
P(V)\approx P_0 \left(1 + \frac{3 V^2}{U_{\mathrm{tip}}^2} \right)+ \frac{P_iv_0}{V} + \frac{1}{2} d_0 \rho s A V^3. \label{eq:PstrApprox}
\end{align}
A typical plot of $P(V)$ versus UAV speed $V$ is shown in Fig.~\ref{F:PowervsSpeed}, together with the three individual power components and the convex approximation given in \eqref{eq:PstrApprox}.

 Two particular UAV speeds that are of high practical interests are the {\it maximum-endurance (ME) speed} and the {\it maximum-range (MR) speed}, which are denoted as $V_{\mathrm{me}}$ and $V_{\mathrm{mr}}$, respectively.

 {\bf ME speed:} By definition, the ME speed $V_{\mathrm{me}}$  is the optimal UAV speed that maximizes the UAV endurance under any given onboard energy $E$. With $E$ given, the UAV endurance with constant speed $V$ is given by $\frac{E}{P(V)}$. Thus, $V_{\me}$ is the optimal UAV speed that minimizes the power consumption, i.e.,
$V_{\me}=\mathrm{arg}~\underset{V\geq 0}{\min}\ P(V)$
Though a closed-form expression for $V_{\me}$ is difficult to  obtain due to the complicated expression of $P(V)$ in \eqref{eq:Pstr}, it can be efficiently found numerically.

{\bf MR speed:} On the other hand, the MR speed $V_{\mr}$ is the optimal UAV speed that maximizes the total traveling distance with any given onboard energy $E$. For any given $E$, the range with constant traveling  speed $V$ can be expressed as $\frac{EV}{P(V)}$. Define the function
\begin{align}
E_0(V)\triangleq \frac{P(V)}{V}&=P_0 \left(\frac{1}{V} + \frac{3 V}{U_{\mathrm{tip}}^2} \right)+ P_i  \left( \sqrt{V^{-4} + \frac{1}{4 v_0^4}}-\frac{1}{2v_0^2}\right)^{1/2} + \frac{1}{2} d_0 \rho s A V^2,\label{eq:fV}
\end{align}
which physically represents the UAV energy consumption per unit travelling distance  in Joule/meter (J/m) with speed $V$. Thus, $V_{\mr}$ can be found as
$V_{\mr}=\mathrm{arg}~\underset{V\geq 0}{\min}\ E_0(V)$
Though a closed-form expression for $V_{\mr}$ is difficult to  obtain, it can be efficiently found numerically. Alternatively, $V_{\mr}$ can also be obtained graphically based on the  power-speed curve $P(V)$, by drawing the tangential line from the origin to the power curve that corresponds to the minimum slope (and hence power/speed ratio) \cite{766}, as illustrated in Fig.~\ref{F:PowervsSpeed}. In practice, we usually have $V_{\me}\leq V_{\mr}\leq V_{\max}$.

With given UAV trajectory $\{\mathbf q(t)\}$, the propulsion energy consumption can be expressed as
\begin{align}
E_1(T_t, \{\mathbf q(t)\})=\int_0^{T_t} P(\|\mathbf v(t)\|)dt,
\end{align}
where
$\mathbf v(t) \triangleq \dot{\mathbf q}(t)$
is the UAV velocity and $\|\mathbf v(t)\|$ is the UAV speed at time instant $t$.

By combining both the communication related energy and the propulsion energy, the total UAV energy consumption can be expressed as
\begin{align}\label{eq:Etotal}
E(T_t, \{\mathbf q(t)\}, \{\lambda_k(t)\})=E_1(T_t, \{\mathbf q(t)\})+P_c \int_0^{T_t} \left(\sum_{k=1}^K \lambda_k(t)\right)dt.
\end{align}

\subsection{Problem Formulation for UAV Energy Minimization}
Our objective is to minimize the UAV total energy consumption, while satisfying the target communication throughput requirement for each of the $K$ GNs. The problem can be formulated as
\begin{align}
\mathrm{(P1):} & \underset{T_t, \{\mathbf q(t)\}, \{\lambda_k(t)\}}{\min}  \   E(T_t, \{\mathbf q(t)\}, \{\lambda_k(t)\}) \notag \\
\mathrm{s.t.} \ & \bar R_k(T_t, \{\mathbf q(t)\}, \{\lambda_k(t)\}) \geq \tilde{Q}_k, \ \forall k\in \mathcal K, \label{eq:RateConstr} \\
& \|\dot {\mathbf q}(t) \| \leq V_{\max}, \ \forall t\in [0, T_{t}], \\
& \mathbf q(0)=\mathbf q_I, \ \mathbf q(T_t)=\mathbf q_F, \label{eq:qFConstr}\\
& \lambda_k(t)\in \{0, 1\}, \ \forall k\in \mathcal K,\  t\in [0, T_t], \label{eq:binaryConstr}\\
& \sum_{k=1}^K \lambda_k(t)\leq 1, \ \forall t\in [0, T_t],
\end{align}
where $\mathbf q_I, \mathbf q_F\in \mathbb{R}^{2\times 1}$ represent the UAV's initial and final locations projected onto the horizontal plane, respectively. Note that depending on practical application scenarios, the constraints on the initial/final UAV locations in \eqref{eq:qFConstr} may or may not be present.

Problem $\mathrm{(P1)}$ requires optimizing the UAV trajectory $\{\mathbf q(t)\}$ and communication scheduling $\{\lambda_k(t)\}$, which are both continuous functions with respect to  time $t$. Therefore, $\mathrm{(P1)}$ essentially involves infinite number of optimization variables. Furthermore, $\mathrm{(P1)}$ includes a complicated cost function for the UAV energy consumption, as well as  non-convex constraints in \eqref{eq:RateConstr} and binary constraints in \eqref{eq:binaryConstr}. Therefore, $\mathrm{(P1)}$ is difficult to be directly solved. In Section~\ref{sec:FlyHoverCommun}, we first consider the simple {\it fly-hover-communicate} protocol to make the problem more tractable, by which $\mathrm{(P1)}$ reduces to a problem with a finite number of optimization variables that only depends on the number of GNs $K$, instead of  the (a priori unknown) mission completion time $T_t$. Then in Section~\ref{sec:Joint}, we propose a general solution to  $\mathrm{(P1)}$ by utilizing the new {\it path disretization} technique to convert it into a discretized equivalent problem with a finite number of  optimization variables, for which at least a locally optimal solution can be found via the SCA technique.


\section{Fly-Hover-Communicate Protocol}\label{sec:FlyHoverCommun}
Fly-hover-communicate is a very intuitive protocol that is also easy to implement in practice. In this protocol, the UAV successively visits $K$ optimized hovering locations, each for one GN, and  communicates with each GN only when it is hovering at the corresponding location. As a result, problem $\mathrm{(P1)}$ reduces to finding the optimal hovering locations and hovering (communication) time allocations for the $K$ GNs, as well as the optimal flying speed and path connecting these hovering locations. In the following, we first consider the special case with only one GN to draw useful insights, and then extend the study to the  general case with multiple GNs.


\subsection{Optimal Fly-Hover-Communicate Scheme for One Single GN}
 For the special case with one single GN, the GN index $k$ is omitted for brevity. Without loss of generality, we assume that the GN is located at the origin with $\mathbf w=\mathbf 0$, and the UAV's initial horizontal location is $\mathbf q_I=[\bar D, 0]^T$.  
 To illustrate the fundamental trade-off between hovering energy and flying energy minimization, we assume that there is no constraint on the UAV's final location in this subsection, where the general case with such a constraint will be studied in Section~\ref{sec:MultiGNs} and Section \ref{sec:Joint}. It is not difficult to see that under such a basic setup, the UAV should only fly along the line segment connecting $\mathbf q_I$ and the GN $\mathbf w$.

 One extreme case of the fly-hover-communicate protocol is that the UAV simply hovers at the initial location $\mathbf q_I$ and communicates with the GN until the aggregated information bits reach the target value $\tilde Q$. However, when the initial horizontal distance $\bar D$ is large, such a strategy usually leads to a very low data rate and hence requires very long mission completion time $T_t$. This in turn leads to high  UAV hovering and communication energy consumption. Alternatively, the UAV could fly closer to the GN and hover at a certain location with a shorter link distance to achieve a higher data rate. This strategy, though requiring additional energy for UAV traveling, reduces the time for data transmission (or hovering), and hence requires less energy for hovering and communication. Therefore, with the fly-hover-communicate protocol, in order to minimize the total UAV energy consumption, there must exist an optimal UAV hovering location that strikes an optimal balance between minimizing the traveling energy and hovering/communication energy.

 Denote by $T_{\tr}$ the UAV traveling time before reaching the hovering location and by $V(t)$ the instantaneous traveling speed towards the GN. Thus, the total traveling distance is  $D_{\tr}=\int_0^{T_{\tr}} V(t) dt$, where we should have  $0\leq D_{\mathrm{tr}}\leq \bar D$.
The total required energy consumption for traveling is
\begin{align}
E_{\tr}(T_{\tr},\{V(t)\})=\int_0^{T_{\tr}}P(V(t))dt. \label{eq:Etr}
\end{align}

Furthermore, based on \eqref{eq:Rkt}, the achievable data rate in bps when the UAV hovers at the point after traveling distance ${D}_{\mathrm{tr}}$ is
\begin{align}
R(D_{\tr})=B\log_2\left(1+ \frac{\gamma_0}{H^2+(\bar D-D_{\mathrm{tr}})^2}\right).
\end{align}
Thus, the required communication time (or equivalently the UAV hovering time $T_{\hov}$) to complete the transmission of $\tilde Q$ bits is
\begin{align}
T_{\hov}=\frac{\tilde Q}{R(D_{\tr})}=\frac{Q}{\log_2\left(1+ \frac{\gamma_0}{H^2+(\bar D-D_{\mathrm{tr}})^2}\right)},
\end{align}
where $Q\triangleq \tilde{Q}/B$ is the bandwidth-normalized throughput requirement in bits/Hz. Thus, the total required UAV hovering and communication energy consumption is
\begin{align}
E_{\hc}(D_{\tr})=(P_h+P_c)T_{\hov}=\frac{(P_h+P_c)Q}{\log_2\left(1+ \frac{\gamma_0}{H^2+(\bar D-D_{\mathrm{tr}})^2}\right)}.
\end{align}
Therefore, the total UAV energy consumption is
\begin{align}
E_{\tot}(T_{\tr}, \{V(t)\}, D_{\tr})=E_{\tr}(T_{\tr}, \{V(t)\})+E_{\hc}(D_{\tr}).
\end{align}
Thus, the energy minimization problem $\mathrm{(P1)}$ reduces to
\begin{align}
\mathrm{(P2)}\ \underset{T_{\tr}, \{V(t)\}, D_{\tr}}{\min} \ &  E_{\tot}(T_{\tr}, \{V(t)\}, D_{\tr})\notag \\
 \text{s.t.} \ & 0\leq V(t) \leq V_{\max}, \forall t\in [0, T_{\tr}], \notag \\
& \int_0^{T_{\tr}} V(t) dt = D_{\tr}, \notag \\
& 0\leq D_{\tr} \leq \bar{D}.
\end{align}
It is not difficult to see that with the fly-hover-communicate protocol, the scheduling variable $\lambda(t)$ in problem $\mathrm{(P1)}$ can be directly determined once the UAV traveling time $T_{\tr}$ and hovering time $T_{\hov}$ are obtained.

\begin{lemma}\label{lemma:OptSpeed}
The optimal solution to problem $\mathrm{(P2)}$ satisfies $T_{\tr}=D_{\tr}/V_{\mr}$ and $V(t)=V_{\mr}$, $\forall t\in [0, D_{\tr}/V_{\mr}]$.
\end{lemma}
\begin{IEEEproof}
Lemma~\ref{lemma:OptSpeed} can be shown by change of variables. The details are omitted for brevity.
\end{IEEEproof}

Lemma~\ref{lemma:OptSpeed} shows that with the fly-hover-communicate protocol, the UAV should travel with a constant speed, which is given by the MR speed, $V_{\mr}$. Let $E_0^\star=E_0(V_{\mr})$ be the minimum UAV energy consumption per unit traveling distance obtained by substituting $V_{\mr}$ into \eqref{eq:fV}. Then problem $\mathrm{(P2)}$ reduces to the following uni-variate optimization problem
\begin{align}
\underset{0\leq D_{\tr} \leq \bar D}{\min} \
D_{\tr}E_0^\star+\frac{(P_h+P_c)Q}{\log_2\left(1+ \frac{\gamma_0}{H^2+(\bar D-D_{\mathrm{tr}})^2}\right)}. \label{eq:PSingleGN2}
\end{align}
 Note that the first term of \eqref{eq:PSingleGN2} increases linearly with the traveling distance $D_{\tr}$, while the second term decreases monotonically  with $D_{\tr}$. Therefore, the optimal solution of $D_{\tr}$ to problem \eqref{eq:PSingleGN2} should balance the energy consumption for traveling and hovering, which can be efficiently obtained via a one-dimensional search.

For the asymptotical case in the low-SNR regime, e.g., $\gamma_0\ll H^2$, by applying the approximation $\ln(1+x)\approx x$ for $|x|\ll 1$, the optimal solution to  \eqref{eq:PSingleGN2} can be obtained in closed-form as
$D_{\tr}^*=\max\left\{0, \bar D -\frac{\gamma_0 E_0^\star}{(2\ln 2) (P_h+P_c)Q}\right\}$.
Define $Q'\triangleq \frac{\gamma_0E_0^\star}{(2\ln 2) (P_h+P_c) \bar D}$. The above result shows that the UAV should move closer (i.e., $D_{\tr}^*>0$) to the GN for communication only when $Q>Q'$. Otherwise, it should simply hover at the initial location $\mathbf q_I$ for communication. Furthermore, the larger the throughput requirement $Q$ is, the closer the UAV should move towards the GT for communication. As $Q$ gets sufficiently large, we have $D_{\tr}^*\rightarrow \bar D$, i.e., the UAV should hover on top of the GN for commutation.


\subsection{Fly-Hover-Communicate for Multiple GNs}\label{sec:MultiGNs}
In this subsection, the fly-hover-communicate protocol is extended to the general case with multiple GNs. In this case, $\mathrm{(P1)}$ reduces to finding the optimal set of $K$ hovering locations, each for communicating with one GN, as well as the traveling path and speed among these hovering locations.

Let $\tilde{\mathbf q}_k\in \mathbb{R}^{2 \times 1}$ denote the horizontal coordinate of the UAV hovering location when it communicates with GN $k$. Based on \eqref{eq:Rkt}, the instantaneous communication rate in bps can be written as
\begin{align}
R_k(\tilde{\mathbf q}_k)=B\log_2\left(1+\frac{\gamma_0}{H^2+\|\tilde{\mathbf q}_k-\mathbf w_k\|^2}\right).
\end{align}
As a result, the total required communication time (or the hovering time at location $\tilde{\mathbf q}_k$) to ensure the target throughput $\tilde{Q}_k$ is given by
\begin{align}
T_k(\tilde{\mathbf q}_k)=\frac{\tilde Q_k}{R_k(\tilde{\mathbf q}_k)}=\frac{Q_k}{\log_2\left(1+ \frac{\gamma_0}{H^2+\|\tilde{\mathbf q}_k -\mathbf w_k\|^2}\right)},
\end{align}
where $Q_k\triangleq \tilde Q_k/B$ is the normalized throughput requirement in bits/Hz. Thus, the total required hovering and communication energy at the $K$ locations is a function of $\{\tilde{\mathbf q}_k\}$, which can be expressed as
\begin{align}
E_{\hc}(\{\tilde{\mathbf q}_k\})& =(P_h+P_c)\sum_{k=1}^K T_k(\tilde{\mathbf q}_k) = \sum_{k=1}^K \frac{(P_h+P_c)Q_k}{\log_2\left(1+ \frac{\gamma_0}{H^2+\|\tilde{\mathbf q}_k -\mathbf w_k\|^2}\right)}.
\end{align}

On the other hand, the total required traveling energy depends on the total traveling distance $D_{\tr}$ to visit all the $K$ hovering locations $\{\tilde{\mathbf q}_k\}$, as well as the traveling speed $V(t)$ among them. Similar to Lemma~\ref{lemma:OptSpeed}, it can be shown that with the fly-hover-communicate protocol, the UAV should always travel with the MR speed $V_{\mr}$. Furthermore, for any given set of hovering locations $\{\tilde{\mathbf q}_k\}$ and initial/final locations ${\mathbf q}_I$ and ${\mathbf q}_F$, the total traveling distance $D_{\tr}$ depends on the visiting order of all the $K$ locations, which can be represented by the permutation variables $\pi(k)\in \{1,\cdots, K\}$. Specifically, $\pi(k)$ gives the index of the $k$th GN served by the UAV. Therefore, we have
\begin{align}
D_{\tr}\left(\{\tilde{\mathbf q}_k\}, \{\pi(k)\}\right)= \sum_{k=0}^{K} \|\tilde{\mathbf q}_{\pi(k+1)}-\tilde{\mathbf q}_{\pi(k)}\|,
\end{align}
where for convenience, we have defined $\tilde{\mathbf q}_{\pi(0)}={\mathbf q}_I$ and $\tilde{\mathbf q}_{\pi(K+1)}={\mathbf q}_F$. Thus, the total required UAV traveling energy with the optimal traveling speed $V_{\mr}$ can be written as
\begin{align}
E_{\tr}\left(\{\tilde{\mathbf q}_k\}, \{\pi(k)\}\right)= E_0^\star D_{\tr}\left(\{\tilde{\mathbf q}_k\}, \{\pi(k)\}\right).
\end{align}
The total UAV energy consumption is thus given by
\begin{align}
E_{\tot}\left(\{\tilde{\mathbf q}_k\}, \{\pi(k)\}\right)=E_{\hc}(\{\tilde{\mathbf q}_k\})+E_{\tr}\left(\{\tilde{\mathbf q}_k\}, \{\pi(k)\}\right).
\end{align}
As a result, the energy minimization problem $\mathrm{(P1)}$ with the fly-hover-communicate protocol reduces to
\begin{align}
\mathrm{(P3)}: \ \underset{\{\tilde{\mathbf q}_k\}, \{\pi(k)\}}{\min} & \  E_{\tot}\left(\{\tilde{\mathbf q}_k\}, \{\pi(k)\}\right)\notag \\
& \text{s.t.}\ \big[ \pi(1), \cdots, \pi(K) \big]\in \mathcal {P},
\end{align}
where $\mathcal P$ represents the set of all the $K!$ possible permutations for the $K$ GNs. Note that with the fly-hover-communicate protocol, the user scheduling parameter $\{\lambda_k(t)\}$ in (P1) can be directly obtained based on the solution to $\mathrm{(P3)}$.

Problem $\mathrm{(P3)}$ is a non-convex optimization problem, whose optimal solution is difficult to  obtain. In fact, even with fixed hovering locations $\{\tilde{\mathbf q}_k\}$, problem $\mathrm{(P3)}$ reduces to the classic TSP \cite{TSP} with pre-determined initial and final locations \cite{957}, which is known to be NP hard. Therefore,  problem $\mathrm{(P3)}$ is also NP hard as it is  more general. Fortunately, by utilizing the existing techniques for solving TSP and applying convex optimization, an efficient approximate solution to $\mathrm{(P3)}$ can be obtained.

 To this end, we first introduce the slack variables  $D_{\tr}$ and $z_k=\|\tilde{\mathbf q}_k-\mathbf w_k\|^2$, by which problem $\mathrm{(P3)}$ can be equivalently written as
\begin{align}
\mathrm{(P3.1)}:  &  \underset{D_{\tr}, \{\tilde{\mathbf q}_k\}, \{\pi(k)\}, \{z_k\}}{\min} \  E_0^\star D_{\tr} +  \sum_{k=1}^K \frac{(P_h+P_c) Q_k}{\log_2\left(1+ \frac{\gamma_0}{H^2+z_k}\right)}\notag \\
 \text{s.t.} \ &  \big[\pi(1), \cdots, \pi(K)\big]\in \mathcal {P}, \label{eq:Constr0} \\
& \sum_{k=0}^{K} \|\tilde{\mathbf q}_{\pi(k+1)}-\tilde{\mathbf q}_{\pi(k)}\| \leq D_{\tr}, \label{eq:Constr1}\\
& \|\tilde{\mathbf q}_k -\mathbf w_k\|^2\leq z_k,\  \forall k\in \mathcal K. \label{eq:Constr2}
\end{align}
Note that at the optimal solution to $\mathrm{(P3.1)}$, all the constraints in \eqref{eq:Constr1} and \eqref{eq:Constr2} must be satisfied with strict equality, since otherwise, we may reduce $D_{\tr}$ or $z_k$ to further reduce the cost function of $\mathrm{(P3.1)}$. 

 Problem $\mathrm{(P3.1)}$ can be interpreted as follows. For each GN $k$, constraint \eqref{eq:Constr2} specifies a disk region centered at the GN location $\mathbf w_k$ with radius $\sqrt{z_k}$. The smaller $z_k$ is, the shorter the communication link distance between the UAV and GN $k$, and hence the less  hovering-and-communication energy required (the second term of the cost function in $\mathrm{(P3.1)}$). However, due to constraint \eqref{eq:Constr1}, this would generally require longer traveling distance $D_{\tr}$ and hence more traveling energy. In fact, for any fixed $\{z_k\}$ such that the  hovering-and-communication energy is fixed, problem $\mathrm{(P3.1)}$ essentially reduces to minimizing the total traveling distance $D_{\tr}$, while ensuring that each of the hovering location $\tilde{\mathbf q}_k$ has a distance no greater than $\sqrt{z_k}$ from the GN $k$. This is essentially the classical traveling salesman problem with neighborhood (TSPN), with pre-determined initial and final locations. TSPN is a generalization of the TSP and hence is NP hard as well, where both the visiting order $\{\pi(k)\}$ and the locations $\{\tilde{\mathbf q}_k\}$ inside the disk region need to be optimized. One effective method for solving TSPN is to firstly ignore the disk radius and solve the TSP problem over the $K$ GN locations $\{\mathbf w_k\}$ to obtain the visiting order $[\hat{\pi}(1),\cdots, \hat{\pi}(K)]$ \cite{957}. Though NP hard, TSP can be approximately solved with high-quality solutions by many existing algorithms \cite{908}. As such, the TSPN then reduces to finding the optimal waypoints $\{\tilde{\mathbf q}_k\}$ with the obtained order $\{\hat{\pi}_k\}$, which is a convex optimization problem and hence can be optimally solved \cite{957}.

 With the above idea, an efficient algorithm is proposed to solve problem $\mathrm{(P3.1)}$. Specifically, the visiting order $\{\pi(k)\}$ is firstly set as  $\{\hat{\pi}(k)\}$ obtained by solving the TSP over the GNs' locations $\{\mathbf w_k\}$. As such, problem $\mathrm{(P3.1)}$ reduces to
 \begin{align}
\mathrm{(P3.2)}:  &  \underset{D_{\tr}, \{\tilde{\mathbf q}_k\}, \{z_k\}, \{\eta_k\}}{\min} \  E_0^\star D_{\tr} +  \sum_{k=1}^K \frac{(P_h+P_c) Q_k}{\eta_k}\notag \\
 \text{s.t.} \ &\sum_{k=0}^K \|\tilde{\mathbf q}_{\hat{\pi}(k+1)}-\tilde{\mathbf q}_{\hat{\pi}(k)}\| \leq D_{\tr}, \label{eq:Constr4}\\
& \|\tilde{\mathbf q}_k -\mathbf w_k\|^2\leq z_k, \ \forall k \in \mathcal K \label{eq:Constr5}\\
& \eta_k \geq 0, \ \forall k\in \mathcal K, \label{eq:etakPositive} \\
& \eta_k \leq \log_2 \left(1+ \frac{\gamma_0}{H^2+z_k}\right), \ \forall k\in \mathcal K, \label{eq:etak}
\end{align}
where we have introduced the slack variables $\{\eta_k\}$. It is noted that the cost function of $\mathrm{(P3.2)}$ and constraints \eqref{eq:Constr4}--\eqref{eq:etakPositive} are all convex. However, the newly introduced constraint \eqref{eq:etak} is non-convex. Fortunately, as the right hand side (RHS) of \eqref{eq:etak} is a convex function, a global lower bound can be obtained based on the first-order Taylor approximation at the local point $z_k^{(l)}$, with the superscript $(l)$ denoting the $l$th iteration:
\begin{align}
\log_2 \left(1+ \frac{\gamma_0}{H^2+z_k}\right)\geq \log_2 \left(1+ \frac{\gamma_0}{H^2+z_k^{(l)}}\right)+\rho_k(z_k-z_k^{(l)}),\label{eq:globalLB}
\end{align}
where $\rho_k=\frac{-\gamma_0 \log_2(e)}{(H^2+z_k^{(l)})(H^2+z_k^{(l)}+\gamma_0)}$.

By replacing the RHS of \eqref{eq:etak} with its lower bound, we have the following problem:
\begin{align}
\mathrm{(P3.3)}:  &  \underset{D_{\tr}, \{\tilde{\mathbf q}_k\}, \{z_k\}, \{\eta_k\}}{\min} \  E_0^\star D_{\tr} +  \sum_{k=1}^K \frac{(P_h+P_c) Q_k}{\eta_k}\notag \\
 \mathrm{s.t.} \ & \eqref{eq:Constr4}-\eqref{eq:etakPositive} , \notag \\
& \eta_k \leq \log_2 \left(1+ \frac{\gamma_0}{H^2+z_k^{(l)}}\right)+\rho_k(z_k-z_k^{(l)}), \ \forall k\in \mathcal K.
\end{align}
It can be verified that problem $\mathrm{(P3.3)}$ is convex, which can thus be efficiently solved by existing convex optimization toolbox such as CVX \cite{227}. Furthermore, due to the global lower bound in \eqref{eq:globalLB}, the optimal value of $\mathrm{(P3.3)}$ provides an upper bound to that of $\mathrm{(P3.2)}$. By successively updating the local point $\{z_k^{(l)}\}$,  the SCA-based algorithm for solving $\mathrm{(P3.2)}$ is summarized in Algorithm~\ref{Algo:SCA1}.

\begin{algorithm}[H]
\caption{SCA-based algorithm for Solving $\mathrm{(P3.2)}$.}\label{Algo:SCA1}
\begin{algorithmic}[1]
\STATE {\bf Initialization}: set the  initial hovering locations $\{\tilde{\mathbf q}_k^{(0)}\}$ and let $z_k^{(0)}=\|\tilde{\mathbf q}_k^{(0)}-\mathbf w_k\|$, $\forall k\in \mathcal K$. Let $l=0$.
\REPEAT
\STATE Solve the convex problem $\mathrm{(P3.2)}$ and denote the optimal solution as $D_{\tr}^*, \{\tilde{\mathbf q}_k^*\}, \{z_k^*\}$, $\{\eta_k^*\}$.
\STATE Update the local point as $z_k^{(l+1)}=z_k^*, \forall k\in \mathcal K$.
\STATE Update $l=l+1$.
\UNTIL the fractional decrease of the objective value of $\mathrm{(P3.2)}$ is below a given threshold $\epsilon$.
\end{algorithmic}
\end{algorithm}

By following similar arguments as in \cite{904} and \cite{768}, it can be shown that Algorithm~\ref{Algo:SCA1} is guaranteed to converge to at least a locally optimal solution to problem $\mathrm{(P3.2)}$ that satisfies the KKT conditions.

 \section{General Solution to $\mathrm{(P1)}$ with Path Discretization and SCA}\label{sec:Joint}
 The fly-hover-communicate protocol in the preceding section gives an efficient solution to $\mathrm{(P1)}$, where the number of optimization variables only depends on $K$, rather than the mission completion time $T_t$. However,  this protocol is strictly sub-optimal since the UAV does not communicate while flying. In this section, we propose a general solution to $\mathrm{(P1)}$ without this assumption via jointly optimizing the UAV trajectory and communication time allocation.

\subsection{Path Discretization}
Problem $\mathrm{(P1)}$ essentially involves an infinite number of  optimization variables coupled in time-continues functions  $\mathbf q(t)$ and $\lambda_k(t)$, as well as the unknown mission completion time $T_t$, thus making it  difficult to be directly solved. To obtain a more tractable form with a finite number of optimization variables, $\mathrm{(P1)}$ can be reformulated by disretizing the variables $\{\mathbf q(t)\}$ and $\{\lambda_k(t)\}$. To this end,  prior works  such as \cite{641} and \cite{904} mostly adopted the method of {\it time discretization}, where the time horizon $[0, T_t]$ is discretized into a finite number of time slots with sufficiently small slot length $\delta_t$. However, this method requires that the UAV mission completion time $T_t$ to be pre-specified, which is not the case for our considered energy minimization problem with $T_t$ being an optimization variable as well. One method to address the above issue is by firstly assuming a certain operation time $T_t$, based on which the time discretization method is applied to solve the corresponding optimization problem, and then exhaustively search for the optimal $T_t$. However, this would require to solve a prohibitively large number of optimization problems, each for a given assumed $T_t$, thus making it impractical especially when the optimal $T_t$ is moderately large. To address this issue, in the following, we propose an alternative discretization method, called {\it path discretization}, with which only one optimization problem needs to be solved.

We first clarify the terminologies of {\it trajectory} versus {\it path}. Generally, a path specifies the route that the UAV follows, i.e., all locations along the UAV trajectory, and it does not involve the time dimension. On the other hand, a trajectory includes  its {\it path} together with the instantaneous travelling speed along the path, and thus it involves the time dimension. With path discretization, the UAV path (instead of time) is discretized into $M+1$ line segments, which are represented by $M+2$ waypoints $\{\mathbf q_m\}_{m=0}^{M+1}$, with $\mathbf q_0=\mathbf q_I$ and $\mathbf q_{M+1}=\mathbf q_F$. 
We impose the following constraints:
\begin{align}
\|\mathbf q_{m+1}-\mathbf q_m\| \leq \Delta_{\max}, \ \forall m, \label{eq:maxDisp}
\end{align}
where $\Delta_{\max}$ is an appropriately chosen value so that within each line segment, the UAV is assumed to fly with a constant velocity and the distance between the UAV and each GN is approximately unchanged. For instance,  $\Delta_{\max}$ could be chosen such that $\Delta_{\max} \ll H$.
Let $T_m$ denote the duration that the UAV remains in the $m$th line segment. The UAV flying velocity along the $m$th line segment is thus given by $\mathbf v_m =\frac{\mathbf q_{m+1}-\mathbf q_m}{T_m}, \ \forall m$.
Furthermore, the total mission completion time $T_t$ is given by $T_t=\sum_{m=0}^{M} T_m$.

As a result, with path discretization, the UAV trajectory $\{\mathbf q(t)\}$ is represented by the $M+2$ waypoints $\{\mathbf q_m\}_{m=0}^{M+1}$, together with the duration $\{T_m\}_{m=0}^{M}$ representing the time that the UAV spends within each line segment. With the given $\Delta_{\max}$, $M$ is chosen to be sufficiently large so that $(M+1) \Delta_{\max}\geq \hat{D}$, where $\hat{D}$ is an upper bound  of the required total UAV flying distance. With such a discretization approach, there is no need to specify the  mission completion time $T_t$ in advance, since it can be directly determined once $\{T_m\}$ are obtained. In addition, to characterize the special hovering status, path discretization only requires two discretization points, i.e., by simply letting $\mathbf q_{m}=\mathbf q_{m+1}$, regardless of the hovering duration $T_m$. This is in a sharp contrast to the existing time discretization approach, where the number of discretization points needs to increase linearly with $T_m$, even when the UAV is hovering and its location remains unchanged.

 As such, the distance between the UAV and each GN $k$ can be written as
\begin{align}
d_{mk}=\sqrt{H^2+\|\mathbf q_m-\mathbf w_k\|^2}, \forall k, m,
\end{align}
where $d_{mk}$ represents the distance between the UAV and GN $k$ when the UAV is at the $m$th line segment along its path. As a result, the corresponding achievable rate expression in \eqref{eq:Rkt} for GN $k$ when the UAV is at the $m$th line segment can be represented as
\begin{align}
R_{mk}=B\log_2\left(1+\frac{\gamma_0}{H^2+\|\mathbf q_m - \mathbf w_k\|^2}\right).
\end{align}
Furthermore, for each line segment $m$ along the UAV path, with TDMA among the $K$ GNs, let $\tau_{mk}\geq 0$ denote the allocated time for the UAV to communicate with GN $k$. Then  constraint \eqref{eq:lambdakt} can be written as $\sum_{k=1}^K \tau_{mk} \leq T_m, \ \forall m\in \{0,\cdots, M\}$.

The aggregated communication throughput for GN $k$ in \eqref{eq:Rbark} can be written as
\begin{align}\label{eq:barRkDisc}
\bar R_k(\{\mathbf q_m\}, \{\tau_{mk}\})=B \sum_{m=0}^M \tau_{mk} \log_2\left(1+\frac{\gamma_0}{H^2+\|\mathbf q_m-\mathbf w_k\|^2}\right).
\end{align}

Furthermore, the UAV energy consumption in \eqref{eq:Etotal} can be written as
\begin{equation}
\small
\begin{aligned}
E&(\{T_m\}, \{\mathbf q_m\},  \{\tau_{mk}\})=\sum_{m=0}^{M} T_m P\left(\frac{\Delta_m}{T_m}\right)+P_c \sum_{m=0}^M \sum_{k=1}^K \tau_{mk} \\
& = P_0  \sum_{m=0}^{M}
\left(T_m + \frac{3 \Delta_m^2}{U_{\mathrm{tip}}^2T_m} \right)+ P_i  \sum_{m=0}^M  \left( \sqrt{T_m^4 + \frac{\Delta_m^4}{4 v_0^4}}-\frac{\Delta_m^2}{2v_0^2}\right)^{1/2} + \frac{1}{2} d_0 \rho s A \sum_{m=0}^M \frac{\Delta_m^3}{T_m^2}+P_c \sum_{m=0}^M \sum_{k=1}^K \tau_{mk},\label{eq:EtotalDiscr}
\end{aligned}
\end{equation}
where $\Delta_m \triangleq \|\mathbf q_{m+1}-\mathbf q_m\|$ is the length of the $m$th line segment. Note that in \eqref{eq:EtotalDiscr}, we have used the expression \eqref{eq:Pstr} and the fact that the UAV speed at the $m$th line segment is $\|\mathbf v_m\|=\Delta_m/T_m$.

As a result, the energy minimization problem $\mathrm{(P1)}$ can be expressed in the discrete form as
\begin{align}
\mathrm{(P4):} & \underset{\{\mathbf q_m\}, \{T_m\}, \{\tau_{mk}\}}{\min}  \  E(\{T_m\}, \{\mathbf q_m\}, \{\tau_{mk}\}) \notag \\
\mathrm{s.t.} \ & 
 \sum_{m=0}^M \tau_{mk} \log_2\left(1+\frac{\gamma_0}{H^2+\|\mathbf q_m-\mathbf w_k\|^2} \right)\geq Q_k, \ \forall k, \label{eq:RateConstrDiscr} \\
& \|\mathbf q_{m+1}- \mathbf q_m\| \leq \min\{\Delta_{\max}, T_m V_{\max}\}, \ \forall m\in \{1,\cdots, M\}, \label{eq:maxSpeedConstr} \\
& \mathbf q_0=\mathbf q_I, \ \mathbf q_{M+1}=\mathbf q_F, \label{eq:qFConstrDiscr}\\
& \sum_{k=1}^K \tau_{mk} \leq T_m, \ \forall m\in \{0,\cdots, M\},\\
& \tau_{mk}\geq 0, \ \forall k\in \mathcal K, m\in \{0,\cdots, M\}. \label{eq:tauConstr}
\end{align}
where \eqref{eq:maxSpeedConstr} corresponds to the maximum UAV speed constraint as well as the maximum segment length constraint.

Notice that in problem $\mathrm{(P4)}$, the constraints \eqref{eq:maxSpeedConstr}--\eqref{eq:tauConstr} are all convex. However, both the cost function $E(\{T_m\}, \{\mathbf q_m\}, \{\tau_{mk}\})$ in \eqref{eq:EtotalDiscr} and the throughput constraint \eqref{eq:RateConstrDiscr} are non-convex. Therefore, problem $\mathrm{(P4)}$ is non-convex and it is difficult to find its globally optimal solution. In the following, we propose an efficient algorithm to find (at least) a locally optimal solution to $\mathrm{(P4)}$ based on the SCA technique.

\subsection{Proposed Solution to $\mathrm{(P4)}$}
Firstly, we deal with the non-convex cost function of $\mathrm{(P4)}$. A closer look at the expression in \eqref{eq:EtotalDiscr} reveals that the first, third, and fourth terms are all convex functions with the respect to $\{\mathbf q_m\}$, $\{T_m\}$, and $\tau_{mk}$, which can be shown by using the fact that perspective operation preserves convexity \cite{202}.  However, the second term is non-convex. To tackle this issue, we introduce slack variables $\{y_m\geq 0\}$ such that
\begin{align}\label{eq:ymsq}
y_m^2 = \sqrt{T_m^4 + \frac{\Delta_m^4}{4 v_0^4}}-\frac{\Delta_m^2}{2v_0^2}, \ \forall m\in \{0,\cdots, M\},
\end{align}
which is equivalent to
\begin{align}
\frac{T_m^4}{y_m^2}= y_m^2 + \frac{\Delta_m^2}{v_0^2}, \ \forall m\in \{0,\cdots, M\}. \label{eq:equalConstrYm}
\end{align}
Therefore, the second term of \eqref{eq:EtotalDiscr} can be replaced by the linear expression $P_i \sum_{m=0}^M y_m$, with the additional constraint \eqref{eq:equalConstrYm}.

On the other hand, to deal with the non-convex constraint \eqref{eq:RateConstrDiscr}, we introduce  slack variables $\{A_{mk}\}$ such that
\begin{align}
A_{mk}^2=\tau_{mk}\log_2 \left(1+\frac{\gamma_0}{H^2+\|\mathbf q_m-\mathbf w_k\|^2}\right).  \label{eq:AmConstr2}
\end{align}
As a result, the constraint \eqref{eq:RateConstrDiscr} can be equivalently written as $\sum_{m=0}^M A_{mk}^2 \geq  Q_k, \ \forall k$.
With the above manipulations, $\mathrm{(P4)}$ can be  written as
\begin{align}
\mathrm{(P4.1):} & \underset{\substack{\{\mathbf q_m\}, \{T_m\}, \{\tau_{mk}\}\\ \{y_m\}, \{A_{mk}\}}}{\min}  \
P_0 \sum_{m=0}^{M}
 \left(T_m + \frac{3 \Delta_m^2}{U_{\mathrm{tip}}^2T_m} \right)+ P_i \sum_{m=0}^M y_m + \frac{1}{2} d_0 \rho s A \sum_{m=0}^M \frac{\Delta_m^3}{T_m^2}+P_c \sum_{m=0}^M \sum_{k=1}^K \tau_{mk}
 \notag \\
\text{s.t.} \ & \sum_{m=0}^M A_{mk}^2 \geq Q_k, \ \forall k, \label{eq:RateConstrDiscr2} \\
& \frac{T_m^4}{y_m^2}\leq y_m^2 + \frac{\|\mathbf q_{m+1}-\mathbf q_m\|^2}{v_0^2}, \ \forall m, \label{eq:equalConstrYm2} \\
& \frac{A_{mk}^2}{\tau_{mk}} \leq \log_2\left(1+\frac{\gamma_0}{H^2+\|\mathbf q_m-\mathbf w_k\|^2} \right), \ \forall m, k, \label{eq:AmConstrIneq} \\
& y_m \geq 0, \forall m, \\
& \eqref{eq:maxSpeedConstr}\text{--}\eqref{eq:tauConstr}. \notag
\end{align}
Note that in $\mathrm{(P4.1)}$, the constraints \eqref{eq:equalConstrYm2} and \eqref{eq:AmConstrIneq} are obtained from  \eqref{eq:equalConstrYm} and \eqref{eq:AmConstr2} by replacing the equality sign with inequality constraints. This does not affect the equivalence between problem $\mathrm{(P4)}$ and $\mathrm{(P4.1)}$. To see this, suppose that at the optimal solution to $\mathrm{(P4.1)}$, if any of the constraint in \eqref{eq:equalConstrYm2} is satisfied with strict inequality, then we may reduce the corresponding value of the slack variable $y_m$ to make the constraint \eqref{eq:equalConstrYm2} satisfied with strict equality, and at the same time reduce the cost function. Therefore, at the optimal solution to  $\mathrm{(P4.1)}$, all constraints in \eqref{eq:equalConstrYm2} must be satisfied with equality. Similarly, there always exists an optimal solution to $\mathrm{(P4.1)}$ that makes all constraints in \eqref{eq:AmConstrIneq} satisfied with equality as well. Thus, problem $\mathrm{(P4)}$ and $\mathrm{(P4.1)}$ are equivalent.

Problem $\mathrm{(P4.1)}$ is still non-convex due to the non-convex constraints \eqref{eq:RateConstrDiscr2}--\eqref{eq:AmConstrIneq}. However, all these three constraints can be effectively handled with the SCA technique by deriving the global lower bounds at a given local point. Specifically, for the constraint  \eqref{eq:RateConstrDiscr2}, it is noted that the left hand side (LHS) is a convex function with respect to $A_{mk}$. By using the fact that the first-order Taylor expansion is a global lower bound of a convex function, we have the following inequality
\begin{align}
A_{mk}^2 \geq {A}_{mk}^{(l)2} + 2 {A}_{mk}^{(l)} (A_{mk}-{A}_{mk}^{(l)}), \label{eq:AmkLB}
\end{align}
where ${A}_{mk}^{(l)}$ is the value of $A_{mk}$ at the $l$th iteration. 

Similarly, for the non-convex constraint \eqref{eq:equalConstrYm2}, the LHS is already a jointly convex function with respect to $y_m$ and $T_m$, and the RHS of the inequality constraint is also a convex function. By applying the first-order Taylor expansion of the RHS, the  following global lower bound can be obtained as
\begin{equation}
\begin{aligned}
y_m^2 & +\frac{\|\mathbf q_{m+1}-\mathbf q_{m}\|^2}{v_0^2}
\geq y_m^{(l)2}+ 2y_m^{(l)}(y_m-y_m^{(l)}) - \frac{\|\mathbf q_{m+1}^{(l)}-\mathbf q_{m}^{(l)}\|^2}{v_0^2}+\frac{2}{v_0^2}(\mathbf q_{m+1}^{(l)}-\mathbf q_{m}^{(l)})^T(\mathbf q_{m+1} - \mathbf q_m), \label{eq:ymLB}
\end{aligned}
\end{equation}
where $y_m^{(l)}$ and $\mathbf q_m^{(l)}$ are the current value of the corresponding variables at the $l$th iteration.

Furthermore, for the non-convex constraint \eqref{eq:AmConstrIneq}, the LHS is already a jointly convex function with respect to $A_{mk}$ and $\tau_{mk}$. In addition, with similar derivation as in \cite{641} and \cite{904}, for any given value $\{{\mathbf q}_m^{(l)}\}$ at the $l$th iteration, a global concave lower bound can be obtained for the RHS of \eqref{eq:AmConstrIneq} as
\begin{align}
\log_2\left( 1+ \frac{\gamma_0}{H^2 +\|\mathbf q_m - \mathbf w_k\|^2}\right)
\geq  R_{mk}^{(l)}(\mathbf q_m), \label{eq:RateLB}
\end{align}
where
\begin{align}
 R_{mk}^{(l)}(\mathbf q_m) = \log_2\left( 1+ \frac{\gamma_0}{H^2 +\|\mathbf {q}^{(l)}_m - \mathbf w_k\|^2}\right)-\beta_{mk} \left(\|\mathbf q_m -\mathbf w_k \|^2 -\|{\mathbf q}^{(l)}_m-\mathbf w_k\|^2 \right),
\end{align}
with
$\beta_{mk}=\frac{(\log_2 e) \gamma_0}{(H^2+\|{\mathbf q}^{(l)}_m -\mathbf w_k \|^2)(\|{\mathbf q}^{(l)}_m -\mathbf w_k \|^2+\gamma_0)}$.

By replacing the non-convex constraints \eqref{eq:RateConstrDiscr2}--\eqref{eq:AmConstrIneq}  of $\mathrm{(P4.1)}$ with their corresponding lower bounds at the $l$th iteration obtained above, we have the following optimization problem:
\begin{align}
\mathrm{(P4.2):} & \underset{\substack{\{\mathbf q_m\}, \{T_m\}, \{\tau_{mk}\}\\ \{y_m\}, \{A_{mk}\}}}{\min}  \
P_0 \sum_{m=0}^{M}
 \left(T_m + \frac{3 \Delta_m^2}{U_{\mathrm{tip}}^2T_m} \right)+ P_i \sum_{m=0}^M y_m + \frac{1}{2} d_0 \rho s A \sum_{m=0}^M \frac{\Delta_m^3}{T_m^2}+P_c \sum_{m=0}^M \sum_{k=1}^K \tau_{mk}
 \notag \\
\text{s.t.} \ & \sum_{m=0}^M \left({A}_{mk}^{(l)2} + 2 {A}_{mk}^{(l)} (A_{mk}-{A}_{mk}^{(l)})\right)\geq Q_k, \ \forall k, \notag \\
& \frac{T_m^4}{y_m^2}\leq  y_m^{(l)2}+ 2y_m^{(l)}(y_m-y_m^{(l)})
 - \frac{\|\mathbf q_{m+1}^{(l)}-\mathbf q_{m}^{(l)}\|^2}{v_0^2}+\frac{2}{v_0^2}(\mathbf q_{m+1}^{(l)}-\mathbf q_{m}^{(l)})^T(\mathbf q_{m+1} - \mathbf q_m), \ \forall m, \notag \\
& \frac{A_{mk}^2}{\tau_{mk}} \leq R_{mk}^{(l)}(\mathbf q_m), \ \forall m, k, \notag  \\
& y_m \geq 0, \forall m, \notag \\
& \eqref{eq:maxSpeedConstr}\text{--}\eqref{eq:tauConstr}. \notag
\end{align}
It can be verified that problem $\mathrm{(P4.2)}$ is a convex optimization problem, which can thus be efficiently solved by using standard convex optimization techniques or existing software toolbox such as CVX. Note that due to the global lower bounds in \eqref{eq:AmkLB}--\eqref{eq:RateLB}, if the constraints of problem $\mathrm{(P4.2)}$ are satisfied, then those for the original problem $\mathrm{(P4.1)}$ are guaranteed to be satisfied as well, but the reverse is not necessarily true. Thus, the feasible region of $\mathrm{(P4.2)}$ is in general a subset of that for $\mathrm{(P4.1)}$, and the optimal value of $\mathrm{(P4.2)}$ provides an upper bound to that of $\mathrm{(P4.1)}$. By successively updating the local point at each iteration via solving $\mathrm{(P4.2)}$, an efficient algorithm is obtained for the non-convex optimization problem $\mathrm{(P4.1)}$ or its original problem $\mathrm{(P4)}$. The algorithm is summarized as Algorithm~\ref{Algo:SCA}.

\begin{algorithm}[H]
\caption{SCA-based algorithm for $\mathrm{(P4)}$.}\label{Algo:SCA}
\begin{algorithmic}[1]
\STATE {\bf Initialization}: obtain a feasible $\{\mathbf q_m^{(0)}\}$, $\{T_m^{(0)}\}$, and $\{\tau_{mk}^{(0)}\}$ to $\mathrm{(P4)}$. Let $l=0$.
\REPEAT
\STATE Calculate the current values $\{y_m^{(l)}\}$ and $\{A_{mk}^{(l)}\}$ based on \eqref{eq:ymsq} and \eqref{eq:AmConstr2}, respectively.
\STATE Solve the convex problem $\mathrm{(P4.2)}$, and denote the optimal solution as $\{\mathbf q_m^*\}$, $\{T_m^*\}$, $\{\tau_{mk}^*\}$.
\STATE Update the local point $\q^{(l+1)}_m=\mathbf q_m^*$, $T_m^{(l+1)}=T_m^*$, and $\tau_{mk}^{(l+1)}=\tau_{mk}^*$.
\STATE Update $l=l+1$.
\UNTIL the fractional decrease of the objective value of $\mathrm{(P4.2)}$ is below a given threshold $\epsilon$.
\end{algorithmic}
\end{algorithm}

By following similar arguments as in \cite{904} and \cite{768}, it can be shown that Algorithm~\ref{Algo:SCA} is guaranteed to converge to at least a locally optimal solution that satisfies the KKT conditions of  problem $\mathrm{(P4.1)}$.

\begin{remark}
While Algorithm~\ref{Algo:SCA} is proposed to minimize the UAV energy consumption, it can be similarly applied for UAV communication with other design metrics, such as the following mission completion time minimization problem, by replacing the cost function of $\mathrm{(P4)}$ with $\sum_{m=0}^M T_m$.
\end{remark}

\section{Numerical Results}\label{sec:Numerical}
This section provides numerical results  to validate the proposed designs. The UAV altitude is set as $H=100$ m and the total communication bandwidth is $B=1$ MHz. The received SNR at the reference distance of 1 m is $\gamma_0=60$ dB. As a result, the maximum received SNR when the UAV is just above each GN is $\gamma_0/H^2=20$ dB. The communication related power consumption at the UAV is fixed as $P_c=50$ W. For the UAV's propulsion power consumption, the corresponding parameters are specified in Table~\ref{table:notations}. 
The maximum flying speed is $V_{\max}=60$ m/s. The UAV's initial and final locations are  set as $\mathbf q_I=[0, 0]^T$ and $\mathbf q_F=[800 \text{ m}, 800 \text{ m}]^T$, respectively. We consider the setup with $K=3$ GNs, with their locations shown in red squares in Fig.~\ref{F:trajectory}. We assume that all GNs have identical throughput requirements, i.e., $\tilde Q_k = \tilde Q$, $\forall k\in \mathcal K$.

\begin{figure}
\centering
\includegraphics[width=0.35\linewidth]{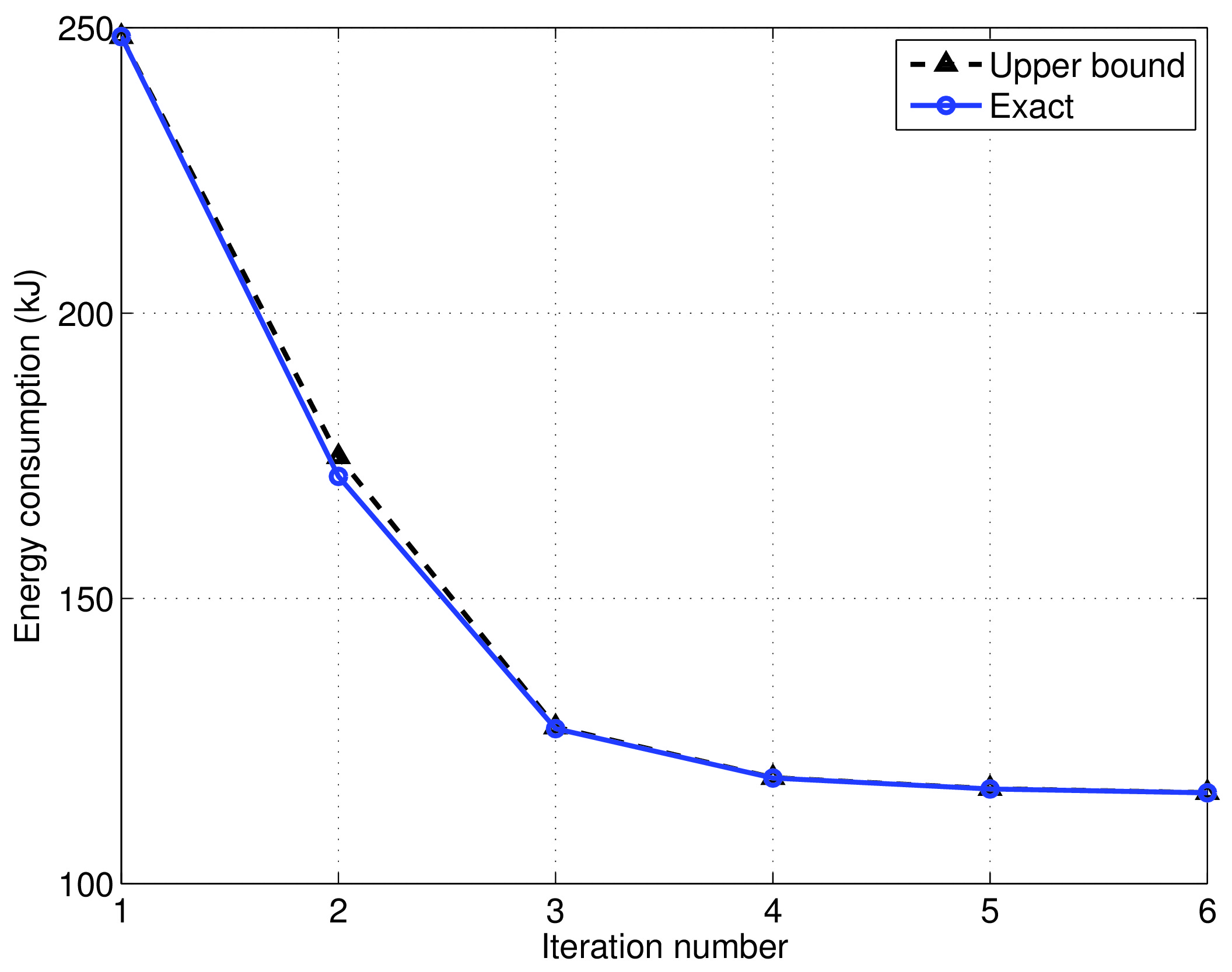}
\caption{Convergence of Algorithm~\ref{Algo:SCA} for UAV energy minimization.\vspace{-3ex}}\label{F:Convergence}
\end{figure}

\begin{figure}
\centering
\begin{subfigure}{0.45\textwidth}
\includegraphics[width=0.8\linewidth]{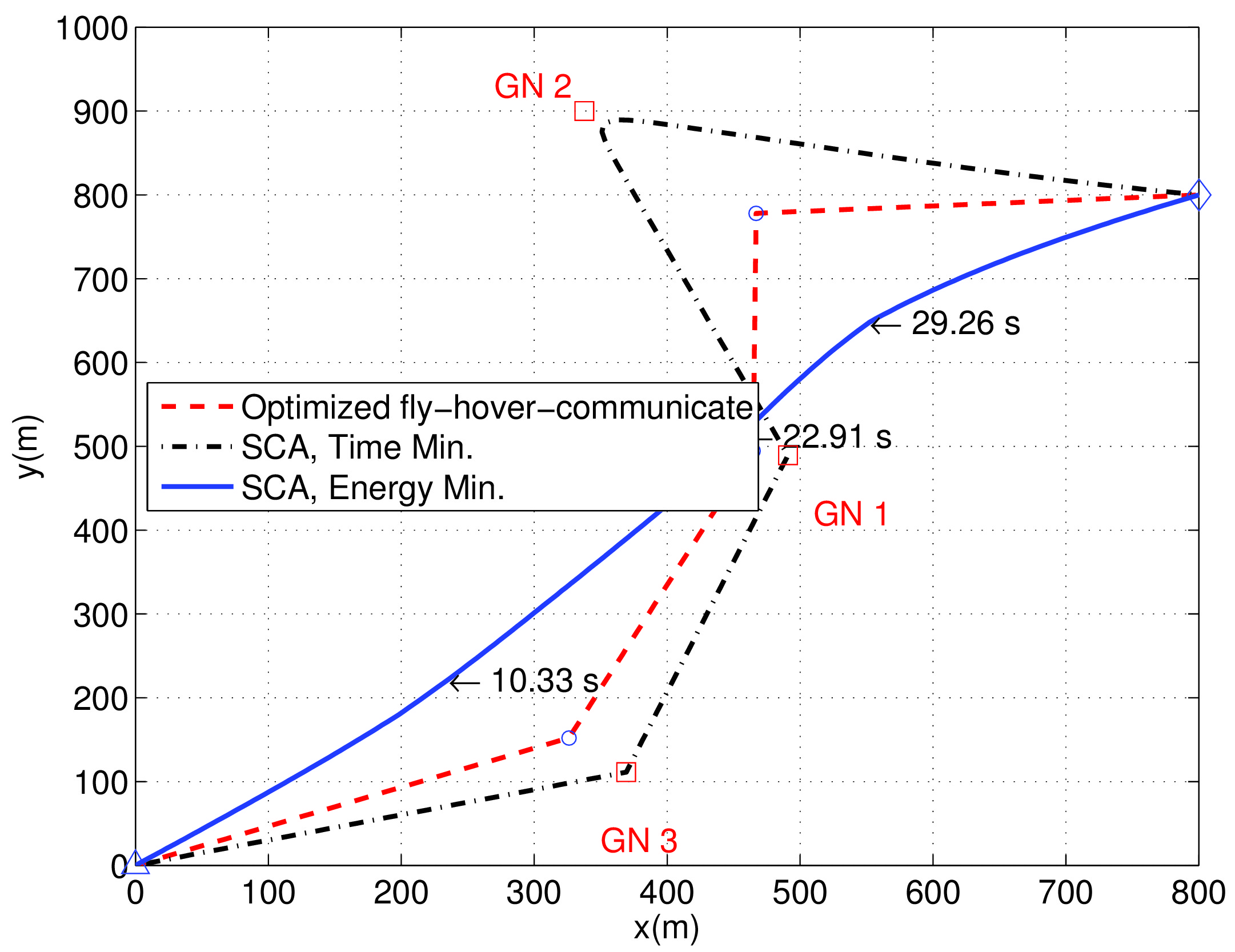}
\caption{$\tilde Q=50$ Mbits.}
\end{subfigure}
\hspace{0.02\textwidth}
\begin{subfigure}{0.45\textwidth}
\includegraphics[width=0.8\linewidth]{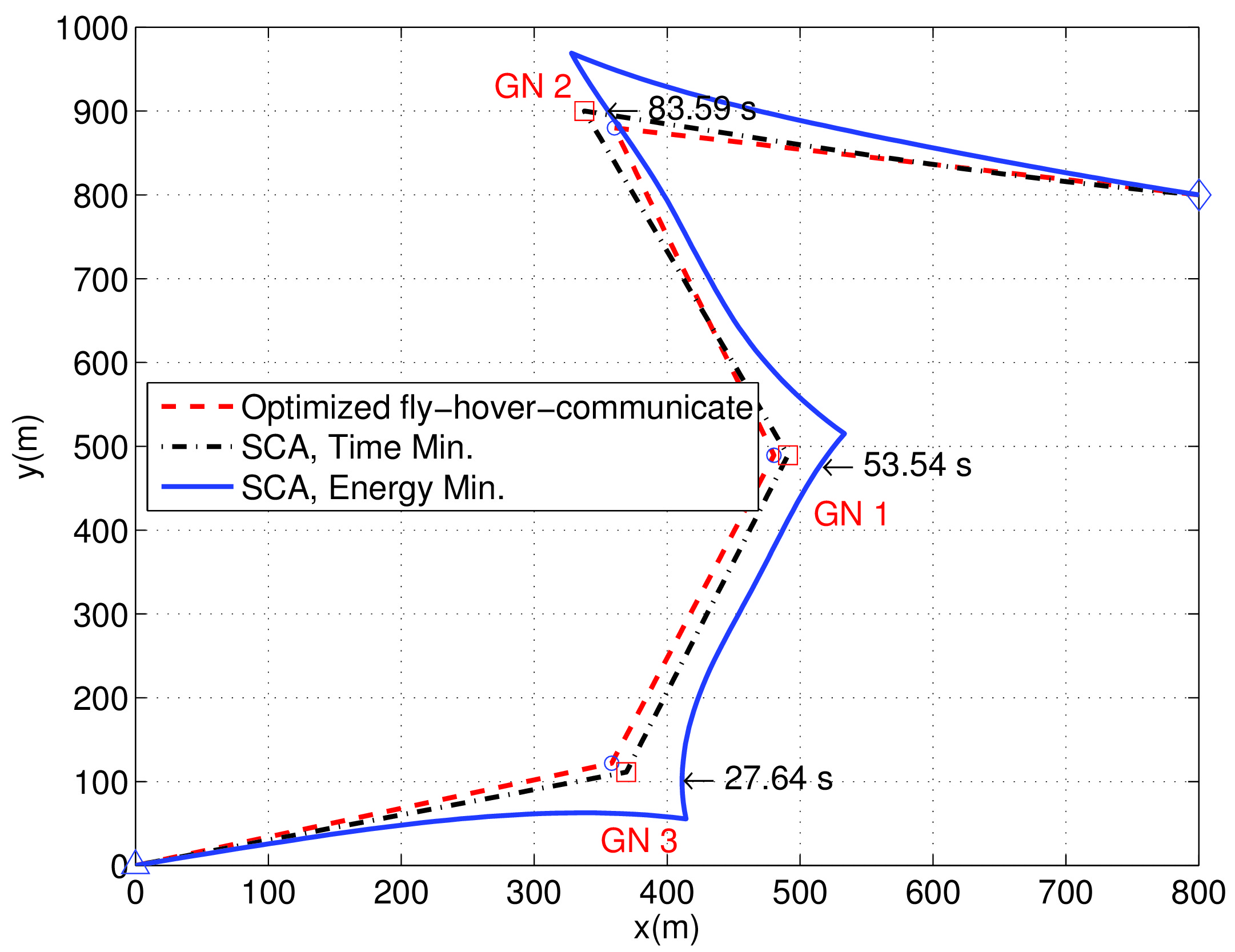}
\caption{$\tilde Q=200$ Mbits.}
\end{subfigure}
\caption{UAV trajectories with three different designs. Red squares denote GNs, and blue circles represent the optimized hovering locations in the fly-hover-communicate protocol.\vspace{-3ex}}\label{F:trajectory}
\end{figure}

First, we study the convergence of Algorithm~\ref{Algo:SCA} (Note that the convergence of Algorithm~\ref{Algo:SCA1} can be shown similarly, for which the result is omitted due to the space limitation). The UAV initial path $\{\mathbf q_m^{(0)}\}$ is set as that obtained by the optimized fly-hover-communicate protocol proposed in Section~\ref{sec:FlyHoverCommun}, and the initial duration $\{T_m^{(0)}\}$ at each line segment and communication time allocation $\{\tau_{mk}^{(0)}\}$ is obtained by letting $T_m^{(0)}=\bar T$, $\forall m$, and $\tau_{mk}^{(0)}=\bar T/K$, $\forall m, k$, where $\bar T$ is the minimum value that makes $\mathrm{(P4)}$ feasible. Fig.~\ref{F:Convergence} shows the convergence of Algorithm~\ref{Algo:SCA} for throughput requirement $\tilde Q=200$ Mbits.  The curve ``Upper bound'' corresponds to the obtained objective value of $\mathrm{(P4.2)}$, while ``Exact'' refers to the true UAV energy consumption value calculated based on \eqref{eq:EtotalDiscr}. It is firstly observed that the two curves match quite well with each other, which demonstrates that the upper bound for UAV energy consumption via solving the convex optimization problem $\mathrm{(P4.2)}$ is practically tight. Furthermore, Fig.~\ref{F:Convergence} shows that the proposed algorithm converges in a few iterations, which demonstrates the effectiveness of SCA for the proposed joint trajectory and communication time allocation design.

\begin{figure}
\centering
\begin{subfigure}{0.45\textwidth}
\includegraphics[width=0.8\linewidth]{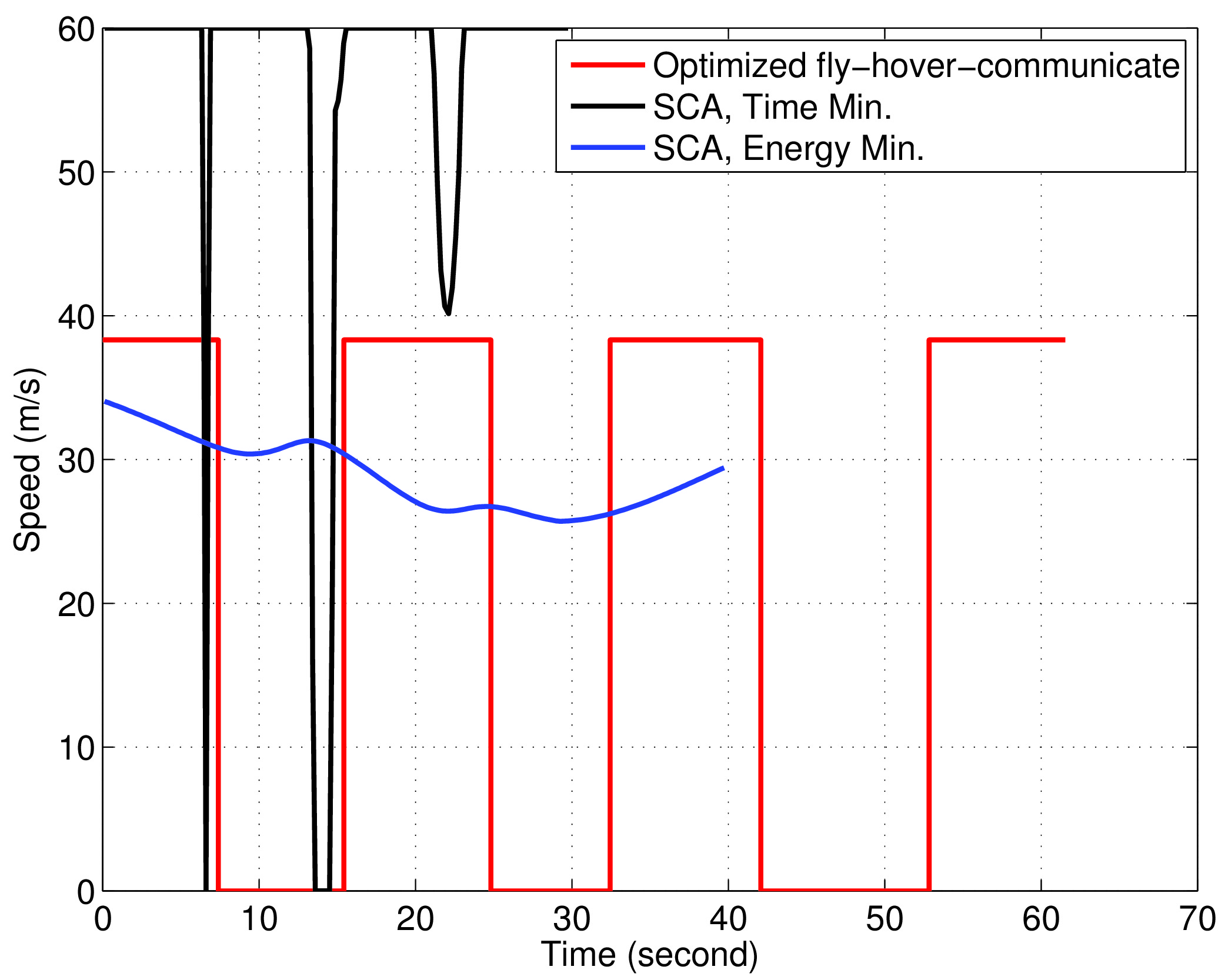}
\caption{$\tilde Q=50$ Mbits.}
\end{subfigure}
\hspace{0.02\textwidth}
\begin{subfigure}{0.45\textwidth}
\includegraphics[width=0.8\linewidth]{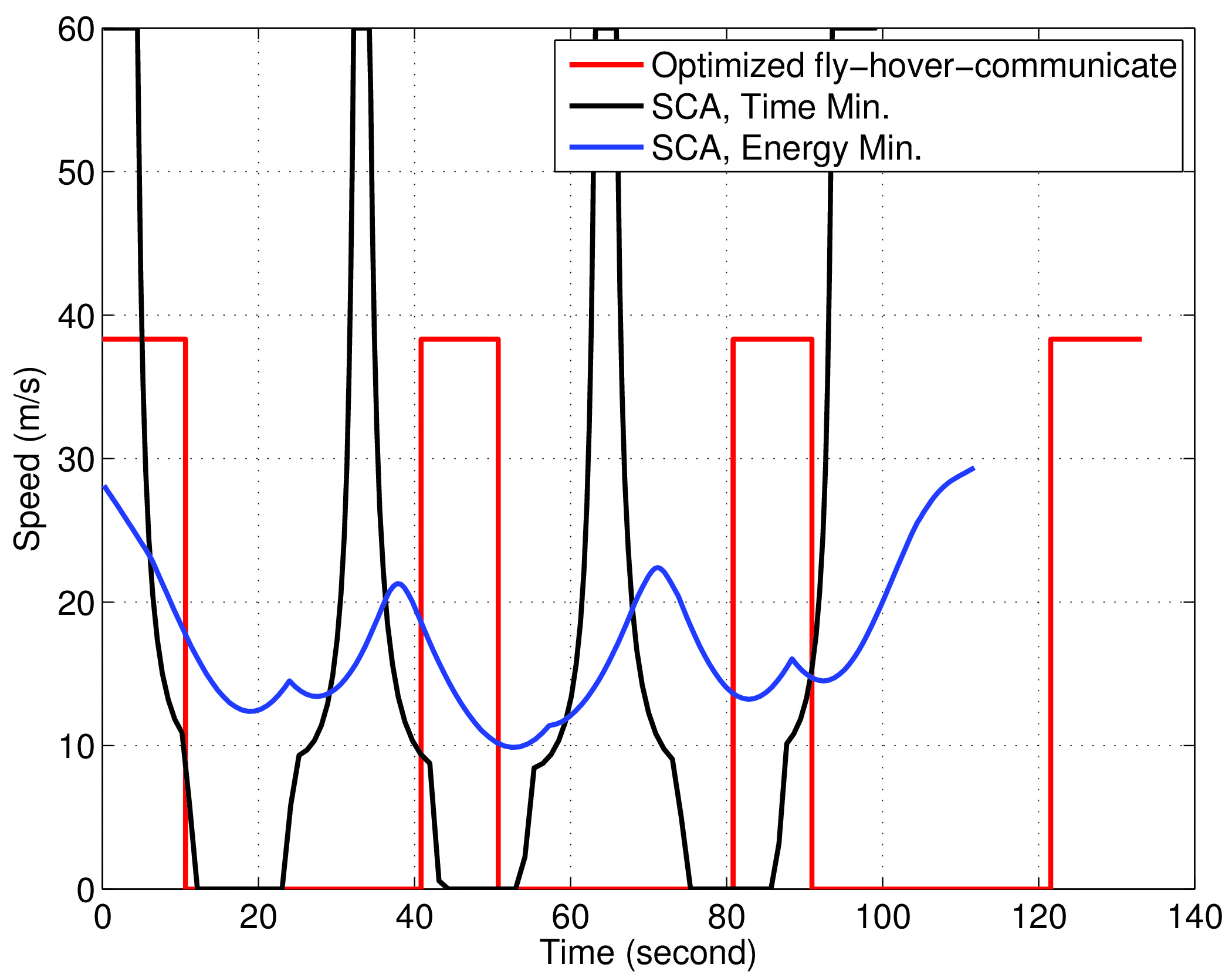}
\caption{$\tilde Q=200$ Mbits.}
\end{subfigure}
\caption{UAV speed versus time for different trajectories in Fig.~\ref{F:trajectory}.\vspace{-3ex}}\label{F:speed}
\end{figure}

For $\tilde Q=50$ Mbits and $200$ Mbits, Fig.~\ref{F:trajectory} and Fig.~\ref{F:speed} respectively show the obtained UAV trajectories and the corresponding UAV speed with three different designs: i) Optimized fly-hover-communicate protocol proposed in Section~\ref{sec:FlyHoverCommun}; ii) The SCA-based {\it energy minimization} design in Algorithm~\ref{Algo:SCA}; and iii) The SCA-based {\it time minimization} design by similarly applying Algorithm~\ref{Algo:SCA}. For the SCA-based energy minimization trajectory, Fig.~\ref{F:trajectory} also shows the corresponding time instant when the UAV reaches the nearest position from each GN, for the convenience of illustrating the corresponding UAV speed shown in Fig.~\ref{F:speed}. It is firstly observed from Fig.~\ref{F:trajectory} that for the proposed fly-hover-communicate protocol, the optimized hovering locations are in general different from the GN locations. This is expected due to the following trade-off: while hovering exactly above each GN achieves the minimal communication link distance and hence reduces the total communication time, it requires the UAV to travel longer distance and hence more energy consumption is needed for UAV flying. With the optimized fly-hover-communicate protocol, a balance between the above two conflicting objectives is achieved via optimizing the hovering locations for communication. By comparing Fig.~\ref{F:trajectory}(a) and Fig.~\ref{F:trajectory}(b), it is observed that the higher the throughput requirement is, the closer the optimized hovering locations will be from the GN locations, as expected. It is further observed from Fig.~\ref{F:speed} that with the optimized fly-hover-communicate protocol, the UAV speed has only two status: flying with the MR speed $V_{\mr}$ between different optimized locations, or hovering above those locations for communicating with the corresponding GN.

For the proposed SCA-based algorithm for energy minimization,  it is found from Fig.~\ref{F:trajectory} that for the case with relatively low throughput requirement of $\tilde Q=50$ Mbits,  the  resulting UAV trajectory is almost a straight flight from $\mathbf q_I$ to $\mathbf q_F$. By contrast, as $\tilde Q$ increases to $200$ Mbits, the UAV needs to deliberately detour its path towards the GNs. Interestingly, it is observed from Fig.~\ref{F:trajectory}(b) and Fig.~\ref{F:speed}(b) that as the UAV approaches the GN, it tends to keep flying around it with a certain speed, instead of hovering directly above it. This is  due to the fact that hovering is not the most power-conserving UAV status, as shown in Fig.~\ref{F:PowervsSpeed}. Therefore, the UAV tends to maintain a certain speed in order to reduce power consumption, though this may make the UAV slightly further away from the GN (thus with smaller instantaneous communication rate). By combining  Fig.~\ref{F:trajectory} and Fig.~\ref{F:speed}, it is found that for the SCA-based trajectory for energy minimization, the UAV will reduce its flying speed when it is close to each GN, as expected.

With the SCA-based algorithm for time minimization, Fig.~\ref{F:trajectory} shows that for both $\tilde Q=50$ Mbits and $\tilde Q=200$ Mbits,  the UAV tends to fly to the top of each GN. This is expected since for time minimization without considering the UAV energy consumption, it is preferable for the UAV to fly at high speed so as to approach the GN as soon as possible to enjoy the favorable communication channel. This is verified by the speed plot in Fig.~\ref{F:speed}.

\begin{figure}
\centering
\includegraphics[width=0.4\linewidth]{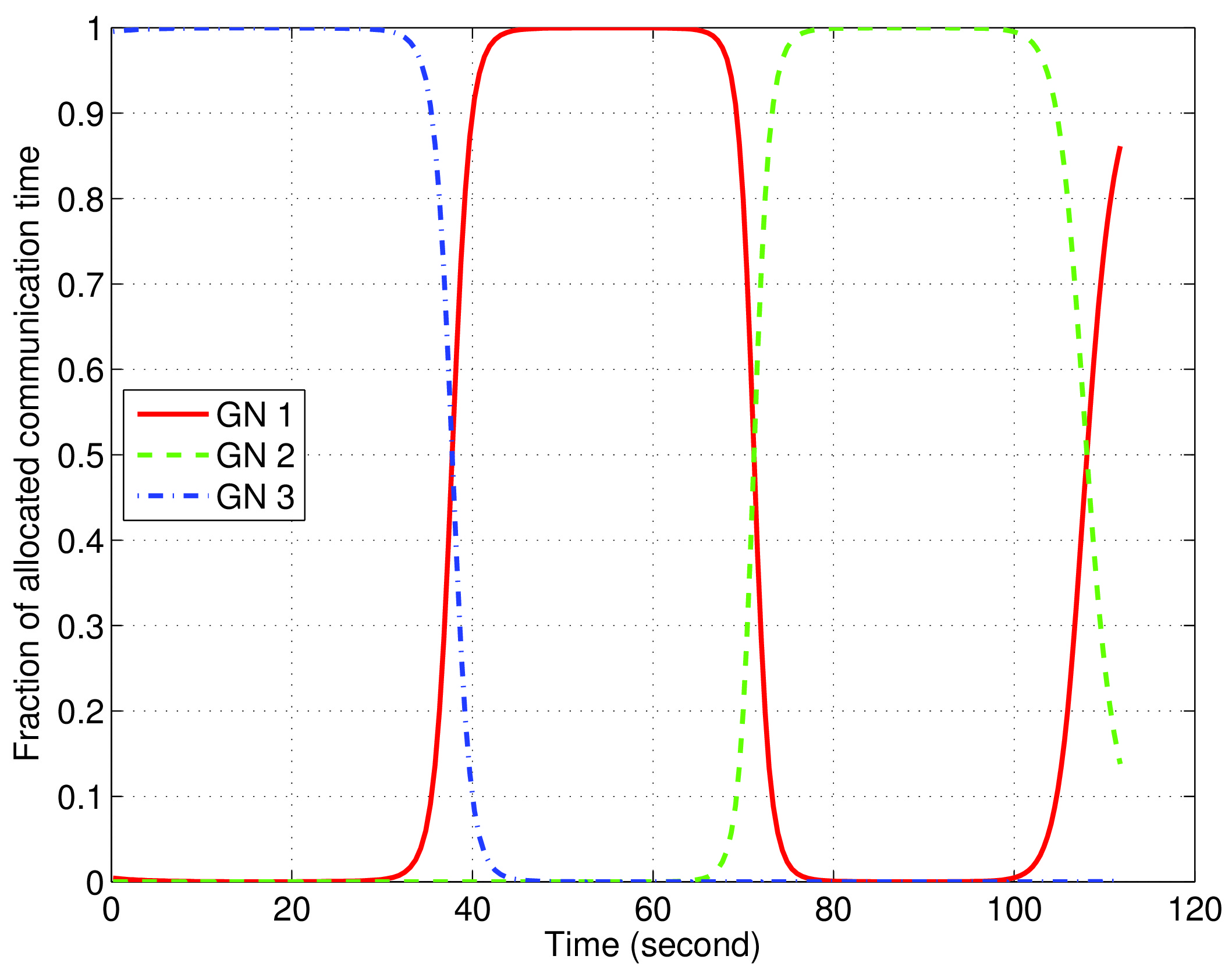}
\caption{Fraction of allocated communicate time for SCA-based energy minimization.\vspace{-3ex}}\label{F:timeAllocation}
\end{figure}

For the proposed SCA-based design for energy minimization with $\tilde Q=200$ Mbits, Fig.~\ref{F:timeAllocation} shows the fraction of the allocated communication time among GNs at each time instant, namely the values of $\tau_{mk}/T_m$. By combining Fig.~\ref{F:trajectory} and Fig.~\ref{F:timeAllocation}, it is found that at each UAV location, more communication time is allocated to the nearer GN, which is expected since allocating resources to better channels in general leads to higher spectrum efficiency.

\begin{figure}
\centering
\begin{subfigure}{0.45\textwidth}
\includegraphics[width=0.9\linewidth]{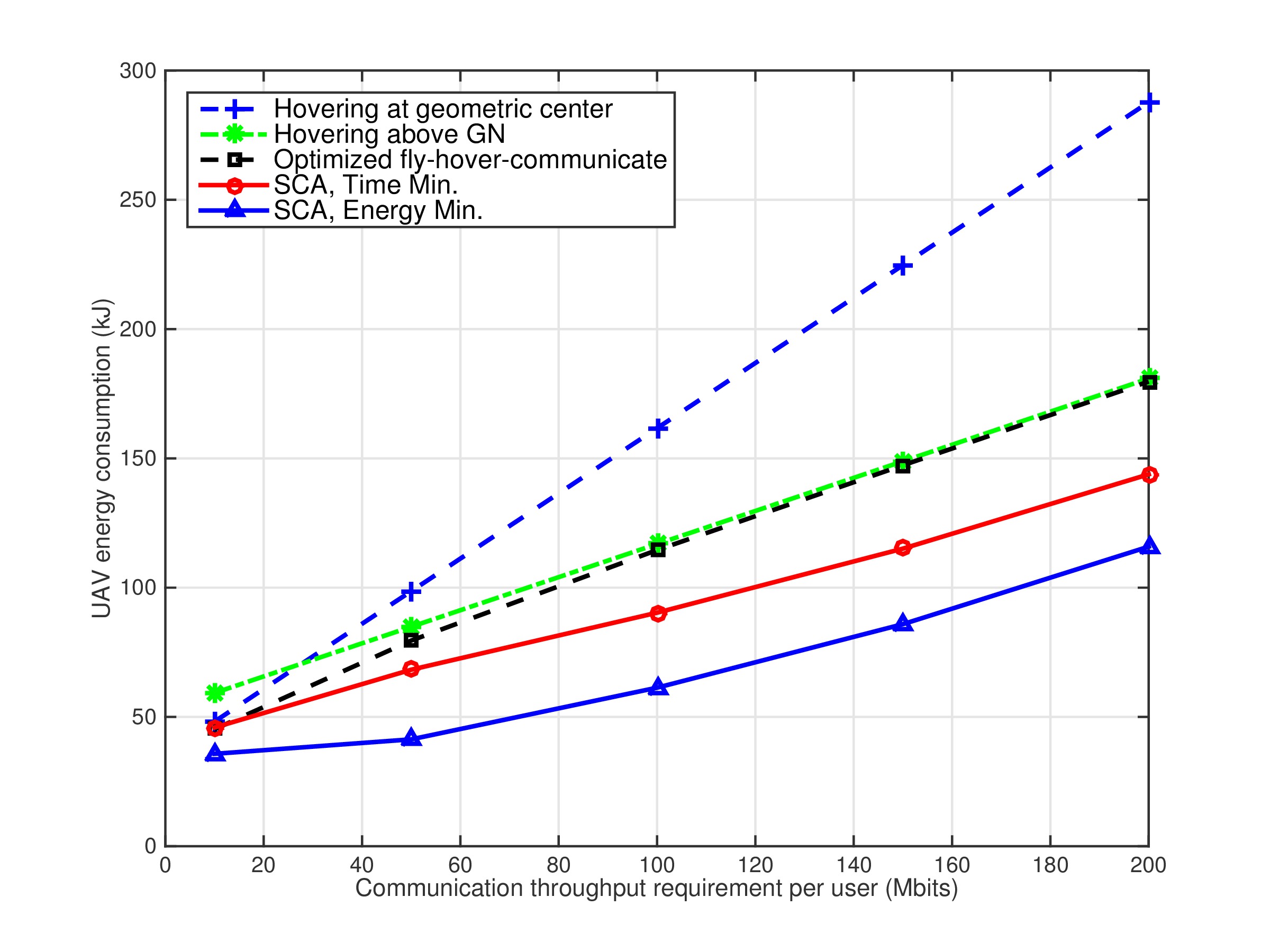}
\end{subfigure}
\hspace{0.02\textwidth}
\begin{subfigure}{0.45\textwidth}
\includegraphics[width=0.9\linewidth]{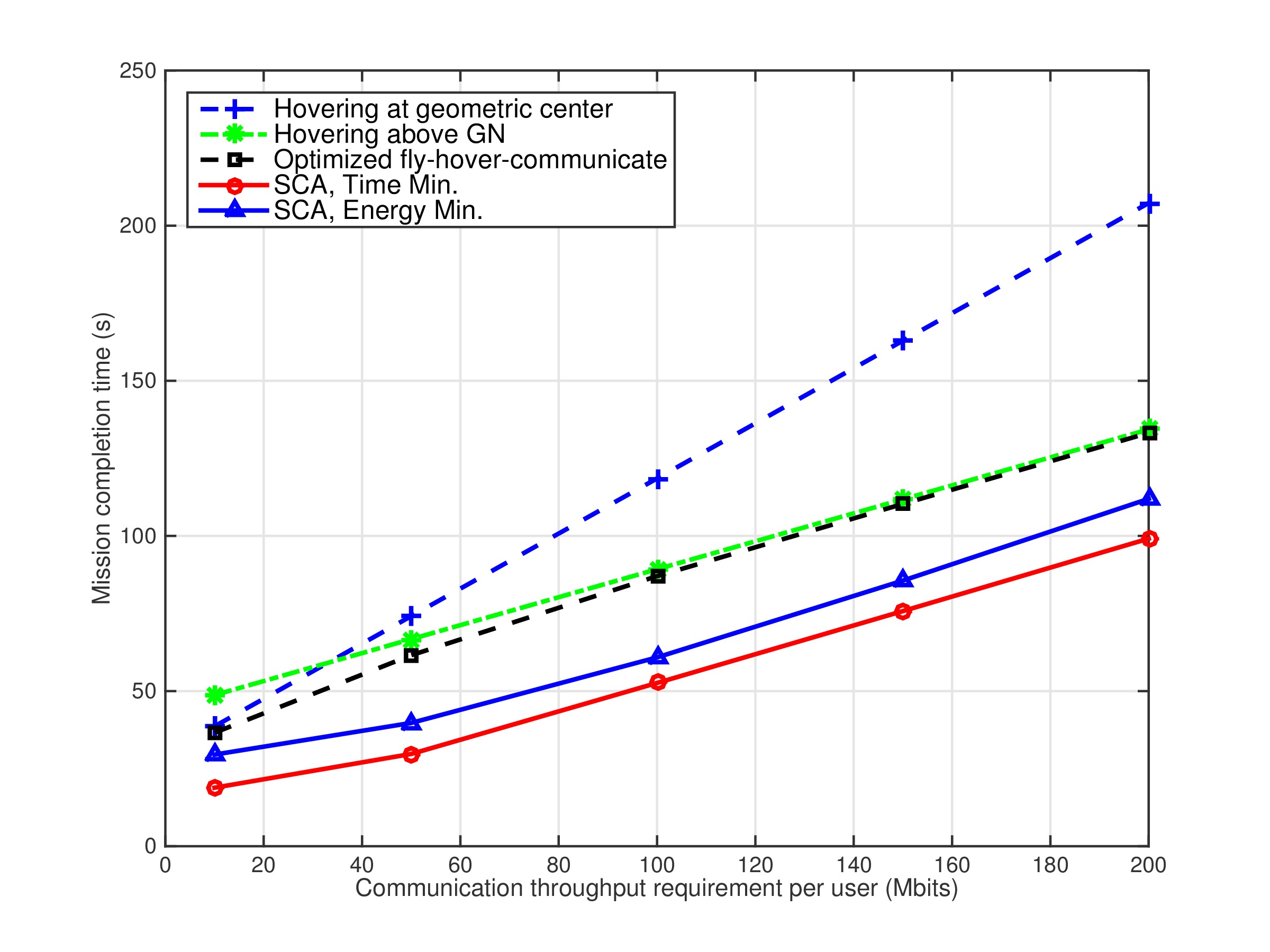}
\end{subfigure}
\caption{Energy consumption and mission completion time versus throughput requirement.\vspace{-3ex}}\label{F:EnergyAndTimeVSThroughput}
\end{figure}

Last, Fig.~\ref{F:EnergyAndTimeVSThroughput} shows the required UAV energy consumption and mission completion time versus the communication throughput requirement $\tilde Q$, respectively. Besides the three designs mentioned above, we also consider two alternative benchmark schemes, namely {\it hovering at geometric center} and {\it hovering above GNs}. Note that these two benchmark schemes correspond to the special cases of the general fly-hover-communicate protocol studied in Section~\ref{sec:FlyHoverCommun}, where the hovering locations are fixed to either the geometric center of the $K$ GNs or each GN, instead of being optimized. It is firstly observed that in terms of both energy consumption and required mission completion time, ``hovering at the geometric center'' outperforms ``hovering above GNs'' for low throughput requirement, whereas the reverse is true as $\tilde Q$ increases. On the other hand, the optimized fly-hover-communicate scheme always outperforms both benchmark schemes, which is expected since it adaptively optimizes the hovering locations according to the communication requirement. Furthermore, with the proposed SCA algorithm either for energy minimization or time minimization, significant performance gains can be achieved. By comparing the two plots in Fig.~\ref{F:EnergyAndTimeVSThroughput}, it is concluded that while minimizing the mission completion time can to certain extent help reduce the energy consumption and vice versa, the two design objectives in general lead to different solutions, and the explicit consideration of UAV energy consumption (instead of via the heuristic time minimization) results in further performance gains in terms of energy saving.

\section{Conclusion}\label{sec:Conclusion}
This paper studies the energy-efficient UAV communication with rotary-wing UAVs. The propulsion power consumption model of rotary-wing UAVs is  derived, based on which an optimization problem is formulated to minimize the total UAV energy consumption, while satisfying the individual target communication throughput requirement for multiple GNs. We first propose an efficient solution based on the simple fly-hover-communicate protocol, which leverages the TSP  and convex optimization techniques to find the optimized hovering locations and durations, as well as the visiting order and speed among these locations. Furthermore, we propose a general solution, with which the UAV  communicates also when flying, by applying a new path discretization approach and the SCA technique. Numerical results show that the proposed designs achieve significant energy saving than other benchmark schemes for rotary-ring UAV enabled wireless communication systems.

\appendices
\section{Power Consumption Model for Rotary-Wing UAVs}\label{A:energyModel}
In this appendix, we derive the power consumption model for rotary-wing UAVs. Note that most of the notations and results follow from the textbook \cite{905}. This appendix is NOT intended to introduce a new physical model for the power consumption of rotary-wing UAVs. Instead, it mainly aims to solicit the existing results in classic aircraft textbooks such as \cite{905} and \cite{766}, to derive an analytical energy model that is suitable for research in UAV communications.
Interested readers may refer to \cite{905} and \cite{766} for more detailed theoretical derivations based on actuator disc theory and blade element theory.
The notations and terminologies used in this appendix are summarized in Table~\ref{table:notations}.

\begin{table}
\footnotesize
\centering
\caption{Notations and terminologies for rotary-wing aircraft.}\label{table:notations}
\begin{tabular}{p{1.3cm}|p{9cm}|p{2.3cm}}
\hline
{\bf Notation}  & {\bf Physical meaning} &{\bf Simulation value} \\
\hline
 $W$ & Aircraft weight in Newton & $100$ \\ 
\hline
$\rho$ & Air density in kg/m$^3$ & $1.225$ \\ 
\hline
 $R$ & Rotor radius in meter (m) & $0.5$ \\
\hline
 $A$ & Rotor disc area in m$^2$, $A\triangleq \pi R^2$ & $0.79$ \\
\hline
$\Omega$ & Blade angular velocity in radians/second & $400$ \\
\hline
$U_{\mathrm{tip}}$ & Tip speed of the rotor blade, $U_{\mathrm{tip}}\triangleq \Omega R$ & $200$ \\
\hline
$b$ & Number of blades & $4$ \\
\hline
$c$ &  Blade or aerofoil chord length & $0.0196$ \\ 
\hline
$s$ & Rotor solidity, defined as the ratio of the total blade area $bcR$ to the disc area $A$, or $s\triangleq \frac{bc}{\pi R}$ & $0.05$ \\
\hline
$S_{FP}$ & Fuselage equivalent flat plate area in $m^2$ & $0.0118$ \\
\hline
$d_0$ & Fuselage drag ratio, defined as $d_0\triangleq \frac{S_{FP}}{sA}$ & $0.3$ \\
\hline
$k$ & Incremental correction factor to induced power & $0.1$ \\
\hline
$T$ & Rotor thrust & -- \\
\hline
$\kappa$& Thrust-to-weight ratio, $\kappa \triangleq \frac{T}{W}$ & -- \\
\hline
$t_c$ & Thrust coefficient based on total blade area, defined as $t_c \triangleq \frac{T}{\rho sA \Omega^2 R^2}$ & -- \\
\hline
$T_D$ & Thrust component along the disc axes. $T_D\approx T$ in practice (Equation (1.39) of \cite{905})& -- \\
\hline
$t_{cD}$ & Thrust coefficient referred to disc axes, $t_{cD}\triangleq \frac{T_D}{\rho sA \Omega^2 R^2}\approx t_c$& -- \\
\hline
$v_0$ & Mean rotor induced velocity in hover, with $v_0=\sqrt{\frac{W}{2\rho A}}$ (see Equation (2.12) of \cite{905} and Equation (12.1) of \cite{766}) & $7.2$ \\
\hline
$v_{i0}$ & Mean rotor induced velocity in forward flight & --\\ 
\hline
$\lambda_i$ & Mean induced velocity normalized by tip speed, $\lambda_i \triangleq \frac{v_{i0}}{\Omega R}$ & --\\
\hline
$\delta$ & Profile drag coefficient. & $0.012$ \\
\hline
$V$ & Aircraft forward speed in m/s & -- \\
\hline
$\hat V$ & Forward speed normalized by tip speed, $\hat V\triangleq \frac{V}{\Omega R}$ & -- \\
\hline
$\alpha_T$ & Tilt angle of the rotor disc, which is small in practice & -- \\
\hline
$\mu$ & Advance ratio, $\mu \approx \hat V =\frac{V}{\Omega R}$ & -- \\
\hline
$q_c$ & Torque coefficient, which, by definition, is directly related to the required power $P$ as $P=q_c \rho s A \Omega^3 R^3$. Note that in many text books, the required rotor power is usually given in terms of $q_c$. & -- \\
\hline
\end{tabular}
\end{table}

For rotary-wing aircrafts in hovering status, the torque coefficient $q_c$ is given by Equation (2.45) of \cite{905}, i.e.,
$q_c=\frac{\delta}{8}+(1+k) \sqrt{\frac{s}{2}}t_c^{3/2}$.
By substituting $t_c=\frac{T}{\rho sA \Omega^2 R^2}$ and noting that the thrust $T$ balances the aircraft weight in hovering status, i.e., $T=W$, we have
\begin{align}
q_c=\frac{\delta}{8} + (1+k)\frac{W^{3/2}}{\sqrt{2} \rho^{3/2}sA^{3/2}\Omega^3R^3}. \label{eq:qcHovering2}
\end{align}
Therefore, by definition of the torque coefficient, the corresponding required power for hovering can be obtained based on the relationship $P=q_c \rho s A \Omega^3 R^3$, which can be expressed as (see also Equation (12.13) of \cite{766})
\begin{align}
P_h=\underbrace{\frac{\delta}{8} \rho s A \Omega^3 R^3}_{\triangleq P_0} + \underbrace{(1+k)\frac{W^{3/2}}{\sqrt{2\rho A}}}_{\triangleq P_i}.\label{eq:Phover}
\end{align}

The derivation of power required for forward flight of a rotary-wing aircraft is much more complicated than that of the fixed-wing counterpart \cite{904}. Fortunately, under some mild assumptions, e.g., the drag coefficient of the blade section is constant, the torque coefficient $q_c$ for an aircraft in forward level flight (zero climbing angle) with speed $V$ is given by Equation (4.20) of \cite{905}, i.e.,
\begin{align}
q_c=\frac{\delta}{8}(1+3 \mu^2) + (1+k) \lambda_i t_{cD} + \frac{1}{2} \hat{V}^3 d_0. \label{eq:qcForward}
\end{align}
 By substituting with $\mu \approx \hat V= \frac{V}{\Omega R}$ and $t_{cD}=\frac{T}{\rho s A \Omega^2 R^2}$, 
   $q_c$ in \eqref{eq:qcForward} can be explicitly written as a function of the forward speed $V$ and rotor thrust $T$ as
\begin{align}
q_c(V, T)=\frac{\delta}{8}\left(1+ \frac{3 V^2}{\Omega^2 R^2}\right) + \frac{(1+k) T \lambda_i }{\rho s A \Omega^2 R^2}  + \frac{1}{2}d_0 \frac{V^3}{\Omega^3 R^3}. \label{eq:qcForward2}
\end{align}
By the definition of the torque coefficient, the required power can be written as a function of $V$ and $T$ as
\begin{align}
P(V, T)&\triangleq q_c(V, T) \rho s A \Omega^3 R^3 \notag \\
&=  P_0 \left(1 + \frac{3 V^2}{\Omega^2 R^2} \right)+(1+k) T v_{i0}  + \frac{1}{2} d_0 \rho s A V^3, \label{eq:Pforward2}
\end{align}
where $v_{i0}=\lambda_i \Omega R$ is the mean induced velocity. Furthermore, based on Equation (3.2) of \cite{905}, for a rotary-wing aircraft with forward speed $V$ and rotor thrust $T$, the mean induced velocity can be calculated as
\begin{align}
v_{i0}& =\left( \sqrt{\frac{T^2}{4\rho^2 A^2}+\frac{V^4}{4}} -\frac{V^2}{2}\right)^{1/2}= v_0 \left( \sqrt{\kappa^2 + \frac{V^4}{4 v_0^4}}-\frac{V^2}{2v_0^2}\right)^{1/2}, \label{eq:vi0}
\end{align}
where $v_0\triangleq \sqrt{\frac{W}{2\rho A}}$ is the mean induced velocity in hover and we have defined $\kappa$ as the thrust-to-weight ratio, i.e., $\kappa\triangleq \frac{T}{W}$. It can be shown that for any given thrust $T$ or $\kappa$, $v_{i0}$ is a decreasing function of $V$. 
By substituting \eqref{eq:vi0} into \eqref{eq:Pforward2}, the required power for forward flight can be more explicitly written as
\begin{align}
P(V, \kappa)
=& \underbrace{P_0 \left(1 + \frac{3 V^2}{\Omega^2 R^2} \right)}_{\text{blade profile}} + \underbrace{P_i \kappa \left( \sqrt{\kappa^2 + \frac{V^4}{4 v_0^4}}-\frac{V^2}{2v_0^2}\right)^{1/2}}_{\text{induced}} + \underbrace{\frac{1}{2} d_0 \rho s A V^3}_{\text{parasite}}, \label{eq:Pforward3}
\end{align}
where $P_0$ and $P_i$ are two constants defined in \eqref{eq:Phover}. 

\begin{figure}
\centering
\includegraphics[scale=0.5]{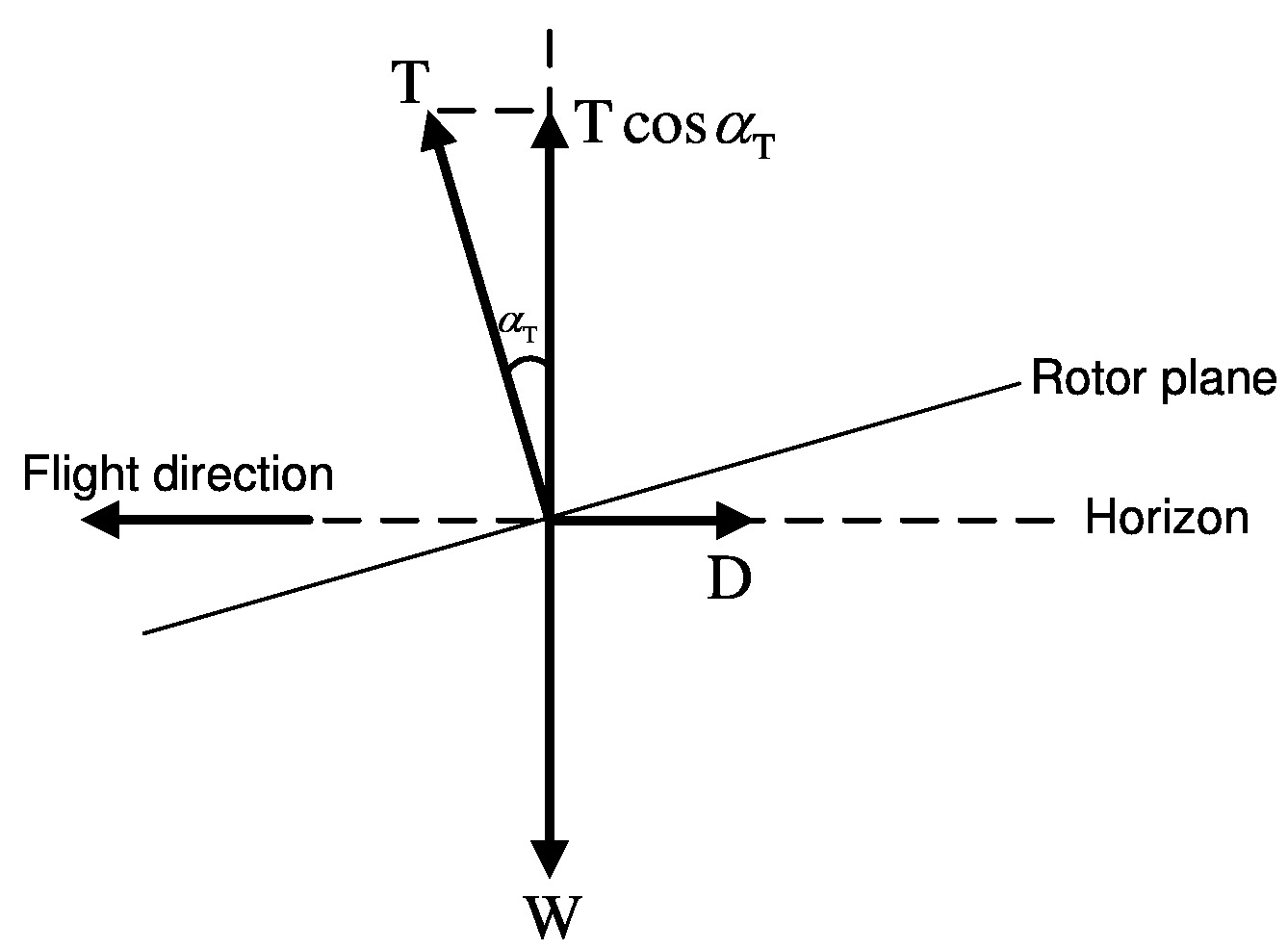}
\caption{Schematics of the main forces acting on the aircraft in straight flight.\vspace{-5ex}}\label{F:Forces}
\end{figure}

To obtain a more explicit expression of the required power  in \eqref{eq:Pforward3}, we need to determine the rotor thrust $T$ or the thrust-to-weight ratio $\kappa$.
Fig.~\ref{F:Forces} shows simplified schematics of the longitudinal forces acting on the aircraft in straight level flight  (see also Figure 13.2 of \cite{766}), which include the following forces: (i) $T$: rotor thrust, normal to the disc plane and directed upward; (ii) $D$: drag of fuselage, which is in the opposite direction of the aircraft velocity; and (iii) $W$: the aircraft weight. Due to the balance of forces in vertical direction, we have $T \cos \alpha_T =W$, where $\alpha_T$ is the tilt angle of the rotor disc. Note that in practice, $\alpha_T$ is usually very small, so we have $T\approx W$ or $\kappa \approx 1$ (see also Equation (4.3) of \cite{905}). As a result, the expression in \eqref{eq:Pforward3} reduces to \eqref{eq:Pstr} shown in Section~\ref{sec:model}.

\bibliographystyle{IEEEtran}
\bibliography{IEEEabrv,IEEEfull}

\end{document}